\documentclass[a4paper,11pt]{article}
\usepackage{jheppub} % for details on the use of the package, please see the JINST-author-manual
\usepackage{lineno}
\usepackage{ragged2e}
\usepackage{csquotes}
\usepackage{cmupint}
\usepackage{graphicx} % Required for inserting images
\usepackage[T1]{fontenc}
\usepackage{amsmath}
\usepackage{graphicx}
\usepackage{latexsym,amssymb,lmodern}
\usepackage{float}
\usepackage[colorinlistoftodos]{todonotes}
\usepackage{hyperref}
\usepackage{xcolor}
\usepackage{soul}
\usepackage{cancel}
\usepackage{appendix}

\usepackage{wrapfig}
\usepackage{braket}
\usepackage{physics}
\usepackage{ulem}
\usepackage{cancel}
\usepackage{comment}
\usepackage{enumitem}
\usepackage{slashed}
\usepackage{mathscinet}
\usepackage{dsfont} 
\usepackage{simpler-wick}
\usepackage{fontawesome5}

\newcommand{\beq}{\begin{equation}}
\newcommand{\eeq}{\end{equation}}
\newcommand{\mn}{m_{\text{min}}}
\newcommand{\m}{M}

\definecolor{blue3}{RGB}{31,119,180}
\definecolor{green3}{RGB}{44,160,44}

\makeatletter
\newcommand{\github}[1]{%
   \href{#1}{\faGithubSquare}%
}
\makeatother

\tikzset{graviton/.style={decorate, decoration={snake, amplitude=.4mm, segment length=1.5mm, pre length=.5mm, post length=.5mm}, double}}

\title{Unitarity, Causality, and Solar System Bounds May Significantly Limit the Use of Gravitational Waves to Test General Relativity}

\author{Alexander Cassem, Mark P.~Hertzberg}

\affiliation{Institute of Cosmology, Department of Physics and Astronomy, Tufts University, Medford, MA 02155, USA}

\emailAdd{alexander.cassem@tufts.edu}
\emailAdd{mark.hertzberg@tufts.edu}

\abstract{
The prospect of detecting/constraining deviations from general relativity by studying gravitational waves (GWs) from merging black holes has been one of the primary motivations of GW interferometers like LIGO/Virgo. Within pure gravity, the only possible way deviations can arise is from the existence of higher order derivative corrections, namely higher powers of the Riemann curvature tensor, in the effective action. Any observational bounds  imply constraints on the corresponding Wilson coefficients. 
   At the level of the action, one can  imagine the coefficients are sufficiently large so as to be in principle detectable.
   However, from the point of view of some fundamental principles, namely causality and unitarity, this is much less clear, as we examine here. We begin by reviewing certain known bounds on these coefficients, which together imply a low cut off on the effective theory. 
   We then consider a possible mechanism to generate such terms, namely in the form of many scalars, minimally coupled to only gravity, that can be integrated out to give these higher order operators. We show that a by product of this is the generation of quantum corrections to Newton's potential, whose observable consequences are already ruled out by solar system tests. We point out that over 7 orders of magnitude of improvement in interferometer sensitivity would be required to avoid such solar system constraints. We also mention further constraints from Hawking radiation from black holes.

}

\begin{document}
\maketitle
\flushbottom

\section{Introduction}\label{introduction section}

Searches for either new effects or corrections that deviate  from general relativity (GR) is an ongoing front in both the experimental and theoretical aspects of gravitation. On the experimental side, we have a century of precision tests of GR in the solar system. However, within the solar system, the spatial curvature is very small. More recently, we have LIGO/Virgo \cite{LIGOScientific:2016aoc} detecting gravitational waves from merging black holes, where the spatial curvature is larger. Currently, all observations are broadly compatible with the predictions of GR. On the theoretical side, much has been devoted to 
understanding the space of possible deviations. Within  effective field theory (EFT), the {\it only} way deviations can appear is through higher derivative operators and their possible appearance within an ultraviolet completion.

Constraining higher derivative operators from LIGO/Virgo observations has grown in interest. In particular, Ref.~\cite{Sennett:2019bpc} has put a bound on the quartic power of Riemann operator of $\Lambda_{4,bd} = 1/(150\:\text{km})$. 
In a more EFT-bootstrap manner, \cite{Caron-Huot:2021rmr,Caron-Huot:2022ugt} has constrained the cubic and quartic power of Riemann curvature operators to high numerical precision using constraints on the coefficients of the operators via causality. Whether or not these two regimes tell us something new about the Wilson coefficients of the higher curvature operators is an open question, and one we intend to begin answering. An analysis of corrections to Newton's potential from parity-preserving operators has been considered in \cite{Melville:2024zjq} with time delay implications.

In Section \ref{effective field theory and bounds section} we lay out the higher derivative operators within the EFT for gravity to discuss the causality bounds on the Wilson coefficients. This includes cubic operators (with coefficients $\Lambda_3$) and quartic operators (with coefficients $\Lambda_4$). 
We generate these higher derivative operators from a set of massive scalars minimally coupled to gravity that can be integrated out. 
%\ac{We only consider integrating out scalars minimally coupled to gravity in order to relate back to solar system bounds. Different types of matter can be included for generality, such as the Standard Model, but do not affect the results here as can be seen.}
We summarize bounds on the set of masses $m_i$ and the number  $N$ of particles that can appear via loop corrections.

In Section \ref{one loop observable section} we compute the 1-loop correction associated with this realization. We show that the 1-loop correction contains two diagrams, plus two counter term diagrams that are necessary in order to renormalize non-local divergences. We then compute the resulting amplitude in the non-relativistic limit for two cases, $mr\ll 1$ and $mr\gg 1$. In the latter limit, we find a quantum correction to Newton's potential that is similar to Yukawa corrections allowing us to test the possible strength of the correction with respect to solar system constraints.  We find that if we are try to saturate, or be near, the bound from LIGO/Virgo of $\Lambda_4\sim 1/(150\:\text{km})$, then we should have already seen deviations from general relativity at the level of our solar system. Alternatively to be within solar system bounds, we need $\Lambda_4\gtrsim 1/(8.6\:\text{km})$, which suppresses the quartic Riemann curvature terms by over 7 orders of magnitude, which is a difficult task for future interferometers.

We also include three appendices. Appendix \ref{feynman rules appendix section} includes the graviton-scalar Feynman rules we used to compute our 1-loop result with, the full form of the 1-loop bubble diagram at order $m^4$, $m^2$, and $m^0$, non-local functions that are useful in computing the effective action that is derived in \ref{effective action appendix section} from the amplitude in \ref{total amplitude section}. Finally in section \ref{discussion on R2 appendix section}, we discuss the relevance, or otherwise, of $R^2$ and $R_{\mu\nu}R^{\mu\nu}$ being included in deriving corrections to Newton's law within the EFT of gravity.

\section{Effective Field Theory and Bounds}\label{effective field theory and bounds section}

\subsection{Building the Effective Field Theory}\label{building the effective field theory section}

The effective field theory action for gravity is an expansion in derivatives. We set $\hbar = c = 1$ with metric signature $(+,-,-,-)$ (and adapting the notation of \cite{Caron-Huot:2022ugt}) we write the expansion as
\begin{align}
    S_{\text{gravity}}  = \int d^4x\sqrt{-g}\biggm(&-\Lambda - \frac{2R}{\kappa^2} + \frac{\alpha_2}{\kappa^2}R^{(2)}%\alpha R^{\mu\nu}R_{\mu\nu} + \beta R^2
    -\frac{1}{3\kappa^2}\left(\alpha_3R^{(3)} + \tilde{\alpha}_3\tilde{R}^{(3)}\right)\nonumber\\
    &+ \frac{1}{2\kappa^2}\left(\alpha_4(R^{(2)})^2+\alpha_4'(\tilde{R}^{(2)})^2 + 2\tilde{\alpha}_4R^{(2)}\tilde{R}^{(2)}\right) + \cdots\biggm)
    \label{the action}
\end{align}
and $\kappa^2 = 32\pi G $ is the bare gravitational coupling, $\Lambda$ is the bare cosmological constant, and the $\alpha_i$'s are coefficients, whose values one wishes to constrain. We have also defined 
\begin{align}
    R^{(2)}& = R_{\mu\nu\alpha\beta}R^{\mu\nu\alpha\beta}, &\tilde{R}^{(2)} = R_{\mu\nu\alpha\beta}\tilde{R}^{\mu\nu\alpha\beta}, && \tilde{R}_{\mu\nu\alpha\beta}\equiv\frac{1}{2}\epsilon_{\mu\nu}{}^{\rho\sigma}R_{\rho\sigma\alpha\beta},\nonumber\\
    R^{(3)} &= R_{\mu\nu}{}^{\rho\sigma}R_{\rho\sigma}{}^{\alpha\beta}R_{\alpha\beta}{}^{\mu\nu},&\tilde{R}^{(3)} = R_{\mu\nu}{}^{\rho\sigma}R_{\rho\sigma}{}^{\alpha\beta}\tilde{R}_{\alpha\beta}^{\mu\nu}.
\end{align}
We have only included terms that involve powers of Riemann, but not terms that are powers of Ricci or Ricci tensor, such as $R^2$ or $R^3$, etc. The reason being is that we are primarily interested in the merger of Kerr black holes, whose gravitational waves have been detected. Such black holes have $R=0$ and $R_{\mu\nu}=0$ as they are vacuum solutions. So we only need to focus on powers of Riemann, as this is non-zero even for vacuum solutions. 
One could consider these as local terms, $R^2,\,R_{\mu\nu}R^{\mu\nu}$, etc, and inquire about their possible corrections to Newton's law.
We comment on this issue in Appendix \ref{discussion on R2 appendix section}.

In 4-dimensional spacetime, as we will work in, the quadratic term, $R^{(2)}$, is unimportant as it can be rewritten as the Gauss-Bonnet term plus the local terms, and so it is only a total derivative and has no effect on the classical equations of motion. 

Hence, the interesting leading order terms start at cubic and quartic order in the Riemann tensor.

\subsection{Constraints from Observed Gravitational Waves}\label{constraints from observed grav waves section}

In Ref.~\cite{Sennett:2019bpc}, bounds are placed on these curvature operators from comparison to LIGO/Virgo data (although in Ref.~\cite{Sennett:2019bpc} the cubic term was assumed small, for reasons we will come to below).
We can express these bounds in terms of a physical scales (inverse lengths) $\Lambda_3$ and $\Lambda_4$ that sets the size of the cubic and quartic powers of the Riemann operator in the action. These are related to our Wilson coefficients as 
\begin{equation}
\frac{\alpha_4}{4} = \frac{1}{\Lambda_4^{6}},\quad\quad \frac{\alpha_3}{3!} = \frac{1}{\Lambda_3^{4}}.
\end{equation}
The observational bound found in \cite{Sennett:2019bpc} is 
\beq
\Lambda^{-1}_4\lesssim\Lambda^{-1}_{4,0}\simeq 150\,\text{km}
\eeq
(i.e., we are using the notation $\Lambda_{4,0}$ as short-handed notation for the inverse scale of 150\,km).
This scale naturally emerges since black holes of mass $\sim 30\,M_{\text{sun}}$ have a Schwarzschild radius of $\approx 90$\,km, which is of this order. Although the cubic terms were not directly included in the analysis, a similar bound is expected for them also, i.e., $\Lambda_3^{-1}\lesssim\Lambda^{-1}_{3,0}\sim 150\,\text{km}$. 

It is important to note that such a bound is stronger than current bounds from the solar system. The reason for this is as follows: even though solar system bounds have more precision, they involve much weaker curvatures. 
Furthermore, if we go to scales below this; for example down to several micro-meters, the Cavendish type tests of gravity are still less stringent, as again the curvatures produced by laboratory masses are extremely tiny. So at the level of classical field theory, these bounds from GWs of merging black holes provide the tightest known direct constraints on the coefficients of these higher dimension operators.

One might wonder if at this scale the theory undergoes strong coupling and changes radically at the scale $L\sim 1/\Lambda_{3,4}$ (which would need to be close to $\sim 150\,\text{km}$ to have relevance to existing observations) to be replaced by some new physics. However, since  we have tested gravity to much smaller scales, which has produced nothing new, this seems to be a big problem. In response to this, in \cite{Sennett:2019bpc}  it is suggested that below this scale there is a ``soft UV completion'' in which the new physics stays hidden and Einstein gravity remains intact. While this is a priori possible, in this work we would like to test this possibility against a concrete UV completion, as we come to shortly.

\subsection{Constraints from Causality}\label{constraints from causality section}

As has been known for the past few years \cite{Camanho:2014apa}, if one only includes
$(\text{Riem})^3$ in the effective action for gravitation, and no other higher dimension operators, then there is superluminality. However, when the quartic operators $(\text{Riem})^4$ are included, then non-zero cubic terms are allowed, so long as an inequality between the cubic and quartic coefficients (which will be explained more in-depth in section \ref{higher derivative operators section}), is satisfied (see \cite{Caron-Huot:2022ugt} and \cite{Bern:2021ppb}). 
The inequality is
\begin{equation}
    \alpha_4+\alpha_4'\geq \delta (\alpha_3^2+\tilde{\alpha}_3^2)\mn^2
    \label{mainCbound}
\end{equation}
where $\mn$ is the mass-gap associated with the mass of the lightest particle that has been integrated out and $\delta$ is an $\mathcal{O}(1)$ prefactor.
This inequality shows that it is not possible to turn on the cubic coupling without including the quartic coupling also.  
This inequality was originally conjectured and argued for in Ref.~\cite{Camanho:2014apa} where it was suggested $\delta=0.25$, and confirmed by Ref.~\cite{Caron-Huot:2021rmr} where the more precise value is said to be $\delta=0.26$, and also by Ref.~\cite{Bern:2021ppb}. 
Furthermore, as is well known, the quartic coefficients need to obey positivity bounds, namely\footnote{Let us also note that the cubic and quartic coefficients cannot be arbitrarily large. As argued in 
 \cite{Caron-Huot:2021rmr}, they are bounded by
\begin{align}
    (\alpha_3^2+\tilde{\alpha}_3^2)\mn^8\leq 24.9\log(\mn/m_{\text{IR}})-27.6\label{c3 inequality}\\
    (\alpha_4 + \alpha_4')\mn^6\leq 12.3\log(\mn/m_{\text{IR}})-13.5\label{c4 inequality}
\end{align}
where $m_{\text{IR}}$ is an infrared scale.}
\begin{equation}
    \alpha_4\geq 0,\quad\quad\alpha_4'\geq 0,\quad\quad 4\alpha_4\alpha_4'\geq\tilde{\alpha}_4^2.
\end{equation}

Together these imply certain bounds on the mass gap $\mn$ and the coefficients within the EFT  $\Lambda_{3,4}$ as we develop in the next section from a particular UV theory.

\subsection{Higher Derivative Operators from Integrating Out Massive Particles}\label{higher derivative operators section}

%\acc{I made two new subsections to differentiate between the two possible models A and B with respect to the comments made by the editor.}

%\subsubsection{EFT to Model the Solar System}

The above higher derivative operators ultimately arise from whatever is the underlying UV completion of gravity. However, since these Wilson coefficients flow with scale, we do not need to know the full UV completion details, such as string theory. Instead, we can imagine that there are many massive, but somewhat light, particles that we integrate out, and we generate these operators in the infrared.
As a concrete example 
%\ac{that allows us to relate back to solar system constraints}, 
we shall focus on integrating out $N$-minimally coupled massive scalars $\phi_i$. 
The starting action is
\begin{align}
    S  = \int d^4x\sqrt{-g}\biggm(&-\Lambda - \frac{2R}{\kappa^2}  +\sum_i\frac{1}{2}(\partial\phi_i)^2-\sum_i\frac{1}{2}m_i^2\phi_i^2
+ \mathcal{L}_{\text{matter}}+\cdots\biggm)
    \label{the action scalar}
\end{align}
where $ \mathcal{L}_{\text{matter}}$ stands for the remaining matter sector, including the standard model. Here the masses of the scalars are denoted $m_i$; their lightest is $\mn$. Often we will focus on the special case in which all these masses are equal, i.e., $\mn=m_1=m_2=\ldots=m_N$, but we will be general for now.

One can integrate out these scalars to build the low energy EFT organized in a derivative expansion. Specifically, the main method is via helicity amplitudes that compute graviton-graviton scattering in the  EFT. 
At one-loop the coefficients of $(\text{Riem})^3$ and $(\text{Riem})^4$ have been determined \cite{Bern:2021ppb}, with values
\begin{align}
    \alpha_3 = \sum_i\frac{1}{(4\pi)^2}\frac{1}{5040}\frac{1}{m_i^2}\left(\frac{\kappa^2}{2}\right),&\:\:\:
    \alpha_4 = \sum_i\frac{1}{(4\pi)^2}\frac{11}{75600}\frac{1}{m_i^4}\left(\frac{\kappa}{2}\right)^2,\nonumber\\
    \alpha_4' = \sum_i\frac{1}{(4\pi)^2}&\frac{1}{75600}\frac{1}{m_i^4}\left(\frac{\kappa}{2}\right)^2. \label{c3 and c4 coefficient from bern for spin 0}
\end{align}
These coefficients have a summation over the index $i=1,\ldots,N$ for each of the particle species that can be in the loop. The coefficients change depending on the spin of the particle in the loop. For our case, we only quote the results for $s = 0$.\footnote{
For the case of non-zero spin, one finds that $\alpha_3=0$ to this order. In this case, one can still proceed, but one should use Eq.~(\ref{c4 inequality}) rather then Eq.~(\ref{mainCbound}), leading to similar results to those found in this section, up to logarithmic correction factors.}
The coefficients $\tilde{\alpha}_4$ and $\tilde{\alpha}_3$ will not be essential for us here. From \eqref{mainCbound}, we can find a relation between $m_i$ and $N$ given the coefficients in \eqref{c3 and c4 coefficient from bern for spin 0} given as
\begin{equation}
    \mn^2\frac{1}{16128}\frac{\kappa^2}{(4\pi)^2}\frac{\left(\sum_im_i^{-2}\right)^2}{\left(\sum_im_i^{-4}\right)}\leq1
    \label{relationship between N and m}
\end{equation}
(where we took $\delta=1/4$ for simplicity). 
As an example, if all the masses are equal, this expression can be restated as
\begin{equation}
    m\leq 160\frac{m_{\text{pl}}}{\sqrt{N}}\label{loose dvali relationship}
\end{equation}
where $m_{\text{pl}}=1/\sqrt{G}$ is the Planck mass. This is a similar relation found previously in \cite{Dvali:2007hz} relating particle species, their mass, and an EFT cutoff. 
By rewriting this bound in terms of the scales $\Lambda_4$ or $\Lambda_3$, this becomes the following pair of inequalities
\begin{align}
    \Lambda_4\geq \left(\frac{75}{11}\right)^{\!1/6} \frac{\left(\sum_i (\mn/m_i)^{2}\right)^{1/3}}{ 
    \left(\sum_i (\mn/m_i)^{4}\right)^{1/3}}\,\,\mn,
    \quad\quad 
     \Lambda_3\geq \left(\frac{15}{4}\right)^{\!1/4} \frac{\left(\sum_i (\mn/m_i)^{2}\right)^{1/4}}{ 
    \left(\sum_i (\mn/m_i)^{4}\right)^{1/4}}\,\,\mn.
     \label{Lambda4 and Lambda3 bounds with m}
\end{align}
The numerical prefactors $(75/11)^{1/6}$ and $(15/4)^{1/4}$ are both approximately 1.4. 

\subsubsection{All Masses Equal}
If there is only one type of massive particle being integrated out ($m=\mn=m_1=\ldots m_N$) this becomes
\beq
\Lambda_3\geq 1.4\,\mn,\quad\quad \Lambda_4\geq 1.4\,\mn.
\label{Lam43}\eeq

  As we discuss in the next subsection, the prefactor in the above expression for $\Lambda_4$ and $\Lambda_3$ is  larger if there are different masses. So the best case scenario for phenomenological purposes then is that all the masses are equal $m=\mn=m_1=\ldots= m_N$. 
   The more general cases of a distribution of masses will be analyzed in the next subsection.
   
   We can readily invert the above logic to bound $\mn$. 
   If we use the observational bound on the EFT of Ref.~\cite{Sennett:2019bpc} of $\Lambda_{4,0}\simeq (150\text{km})^{-1}$ as a useful reference, then we obtain a bound on the mass of the lightest particle of 
   \beq
   m\leq m_\text{min}= 9.5\times10^{-13}\,\text{eV}(\Lambda_4/\Lambda_{4,0})
\label{massbd}   \eeq
   with the bound becoming even stronger for the lightest mass if there are several different masses. Again for the case in which all masses are equal, we can substitute this bound back into $\alpha_4$, and rewrite in terms of $\Lambda_4$ to obtain
   \beq
  \Lambda_4 = \frac{160\,\text{eV}\,(\Lambda_4/\Lambda_{4,0})^{2/3}}{ N^{1/6}}(m/m_c)^{2/3}.
  \eeq
  Re-arranging and solving for $N$ we obtain
    \beq
N=4\times 10^{84}\,(\Lambda_{4,0}/\Lambda_4)^2(m/m_c)^4
\label{causality bound on N 1}
\eeq
  We can interpret this as a bound on the number of species as
  \beq
N \leq N_c=4\times 10^{84}\,(\Lambda_{4,0}/\Lambda_4)^2
\label{causality bound on N}
\eeq
with equality holding when we saturate the theoretical bound $m=m_c$. Although this is an extremely large $N$, it is important to note that such species are assumed to be in a hidden sector and not directly coupled to the Standard Model at tree level. Their only direct couplings to the Standard Model occur via quantum loops as we now explore and learn their possible observational consequences. One might expect that such a huge number of particles would already have been easily visible at colliders, for example from $e^+e^-$ annihilation into these scalars via the graviton in the s-channel; but it is important to recall that the proposal that is presented in Ref.~\cite{Sennett:2019bpc} is that of the ``soft UV completion", meaning that one should only use the theory at very low energies.

\subsubsection{Distribution of Masses}

%\acc{This is all new and my attempt to satisfy the query made by the editor.}

If there are species of particles with different masses, then it is simple to see that the factor involving sums of powers of masses obeys
\beq
P\equiv\frac{\left(\sum_i (\mn/m_i)^{2}\right)}{ 
    \left(\sum_i (\mn/m_i)^{4}\right)}\geq 1
    \eeq
    since there is a fourth power in the denominator and only a second power in the numerator of terms that are each $\mn/m_i\leq 1$ from  the definition of $\mn$. Here
   equality only occurs when all the masses are equal, but otherwise this factor $P$ is larger than unity. In the latter case, the above inequalities (\ref{Lambda4 and Lambda3 bounds with m}) are even sharper, i.e., the energy scales controlling the higher order operators in the EFT must be even larger than $1.4\,\mn$ the lightest mass being integrated out. Hence if the masses being integrated out are somewhat heavy, then these scales are correspondingly microscopic and one cannot use a large $N$ to alter this conclusion.

Let us now begin an exploration into this more general case of a distribution of masses, rather than the above special case of $m = m_\text{min} = m_1 = \dots = m_N$. 

Let us define a function $\rho(m)$, the density of particle masses per unit mass. The previous case in which all particles have the same mass corresponds to a delta-function
\beq
\rho_d(m)=N\,\delta(m-\mn).\,\,\,\,\,\mbox{(Delta)}
\eeq
However, we now consider a spread out distribution. For this purpose let us consider the following pair of representative examples:
\begin{eqnarray}
&&\rho_p(m)={N\,\mn\over m^2},\,\,\,m\ge\mn,\,\,\,\,\,\mbox{(Power Law)}\label{PL}\\
&&\rho_e(m)={N\over f}\exp\left(-(m-\mn)/f\right),\,\,\,m\ge\mn,\,\,\,\,\,\mbox{(Exponential)}\label{Exp}
\end{eqnarray}
with $\rho=0$ for $m<\mn$ in both cases. Note that we have normalized the distribution such that $\int dm\,\rho(m)=N$ in all cases. In the exponential distribution, the parameter $f$ controls how rapidly the exponental falls off at high masses. For small $f\ll\mn$, this approaches a delta-function. While for large $f\gg\mn$, the distribution is very broad.

Using the above distributions, the above prefactor $P$ can be evaluated. For the fixed mass case (delta function distribution), one has $P_d=1$, as remarked above. However, for these more general distributions, we obtain $P>1$. For smooth distributions, we define $P$ as
\beq
P={\int dm\:\rho(m) (\mn/m)^2\over\int dm\:\rho(m)(\mn/m)^4}
\eeq
The values are  found to be
\begin{eqnarray}
&&P_p={5\over3},\,\,\,\,\,\mbox{(Power Law)}\\
&&P_e=P_e(\mn/f),\,\,\,\,\,\mbox{(Exponential)}
\end{eqnarray}
where $P_e$ is a dimensionless function of the ratio $\mn/f$; its explicit form is given in Appendix \ref{AppDistribution}. The function $P_e$ is a monotonic function of its argument, with limiting values:
\begin{eqnarray}
&&P_e\to 1,\,\,\,\,f\ll\mn\\
&&P_e\to 3,\,\,\,\,f\gg\mn
\end{eqnarray}
This factor can then be inserted into Eq.~(\ref{Lambda4 and Lambda3 bounds with m}). It is then raised to the power of $1/3$ for $\Lambda_4$ or the power of $1/4$ for $\Lambda_3$. This implies the corrections are only moderate. In particular we now have the bounds
\begin{eqnarray}
&&\Lambda_4\geq 1.7\,\mn,\,\,\,\,\,\, \Lambda_3\geq 1.6\,\mn\,\,\,\,\,\,(\mbox{Power Law})\\
&&\Lambda_4\geq 2.0\,\mn,\,\,\,\,\,\, 
\Lambda_3\geq 1.8\,\mn\,\,\,\,\,\,(\mbox{Exponential with}\,\, f\gg\mn)
\end{eqnarray}
In the exponential case we have reported on the $f\gg\mn$ limit. In the opposite limit, $f\ll\mn$ the distribution is that of a delta-function, so it just recovers the 1.4 prefactors of Eq.~(\ref{Lam43}).

%Since the above distributions

%In the general scenario, we can have different number of scalar species that are each minimally coupled to gravity. We would then need to consider the more general bounds on $\Lambda_4$ and $\Lambda_3$ given in \eqref{Lambda4 and Lambda3 bounds with m}. However, this would lead to a model-tuning necessity in order to find a realistically observable signal as can be seen later on in section \ref{comparison to solar system bounds section}. Therefore, we take the case of when $m = m_\text{min} = m_1 = \dots = m_N$ throughout the entire paper, and only comment on the more general case when it is warranted to do so.

\section{One-Loop Observable Consequences}\label{one loop observable section}

\subsection{Massive Scalar Amplitude}\label{massive scalar amplitude section}

In the presence of many light scalars, there can be sizable corrections to the gravitational potential between massive objects.
The setup for computing such corrections to Newton's potential was given in \cite{Donoghue:1994dn} initially with the finalized 1-loop correction in pure gravity given in \cite{Bjerrum-Bohr:2002gqz}. We give the basic setup for the EFT treatment of gravity and leave the further details to the references.

There have been many other studies on loop corrections to physical observables such as time-delay effects from violations of causality for example in \cite{Hollowood:2015elj, deRham:2020zyh, Chen:2021bvg, Caron-Huot:2022jli, deRham:2022gfe, Bhat:2023puy, Caron-Huot:2024tsk, AccettulliHuber:2020oou,AccettulliHuber:2020dal}. However, our focus is on corrections to Newton's potential. And furthermore, we shall determine its observational consequences for the solar system.

The main idea is to treat general relativity (GR) as any other quantum field theory that is non-renormalizable. The standard way to proceed is to expand around a background. Since we will be interested in point sources in asymptotic flat space, it is convenient to expand around Minkowski space as
\begin{equation}
g_{\mu\nu} = \eta_{\mu\nu} + \kappa\, h_{\mu\nu},\label{metric expansion}
\end{equation}
where $\eta_{\mu\nu}$ is the Minkowski metric and $h_{\mu\nu}$ is the graviton field. By inserting into the Einstein-Hilbert action with the scalars and matter (\ref{the action scalar}), one can systematically derive the Feynman rules we will be using in this paper; we summarize these in Appendix \ref{Feynman Rules}. For brevity, we do not include the ghost diagrams from the path integral quantization of this gauge theory, as they will not play a direct role to the order we are working. The ghost diagram contributions have previously been computed in \cite{Bjerrum-Bohr:2002gqz}.

The setup for calculating the 1-loop correction to the potential between two point particles exchanging a graviton with a massive scalar at 1-loop is relatively straightforward.
 The difficulty is solely on the number of terms that arise, which is where we utilize \textit{Mathematica} \cite{Mathematica} and specifically the packages \textit{FeynCalc}, \textit{Package-X}, and \textit{FeynGrav} \cite{Mertig:1990an,Shtabovenko:2020gxv, Shtabovenko:2016sxi,Latosh:2023zsi}. 
 We utilize dimensional regularization with the convention $d = 4-2\epsilon$, where $d$ is the number of spacetime dimensions (we then take $\epsilon\to0$ or $d\to4$ at the end of the calculation). We also use Poincare symmetry to simplify integrals with integrands proportional to $p^\mu p^\nu p^\alpha p^\beta$ and $p^\mu p^\nu$ to integrals with integrands proportional to $p^4$ and $p^2$ times factors of the Minkowski metric $\eta^{\mu\nu}$.

We compute $2\rightarrow 2$ scattering of $N$-pairs of point masses $\m_1$ and $\m_2$ scalars up to 1-loop level in the $t$-channel, focusing on the light scalars $\phi_i$ running in the loop. One can also have gravitons or photons running in the loop, but since we will allow the number of hypothetical new scalars $N$ to be large, the scalar contribution will dominate.

There are four diagrams we need to take into account which can be seen in equations \eqref{bubble loop total feynman diagram} and \eqref{shoestring total feynman diagram} while the third and fourth will be discussed later. The general structure of the loop amplitude is found in equation \eqref{general amplitude structure for diagrams} with the respective components of the matrix elements for equations \eqref{bubble loop total feynman diagram} and \eqref{shoestring total feynman diagram} are given in equations \eqref{matrix element 1pi for bubble diagam} and \eqref{matrix element 1pi for shoestring} respectively.

\begin{equation}
    \vcenter{\hbox{$i\mathcal{M}_1$}} \quad = \quad \vcenter{\hbox{\includegraphics[width=0.35\textwidth]{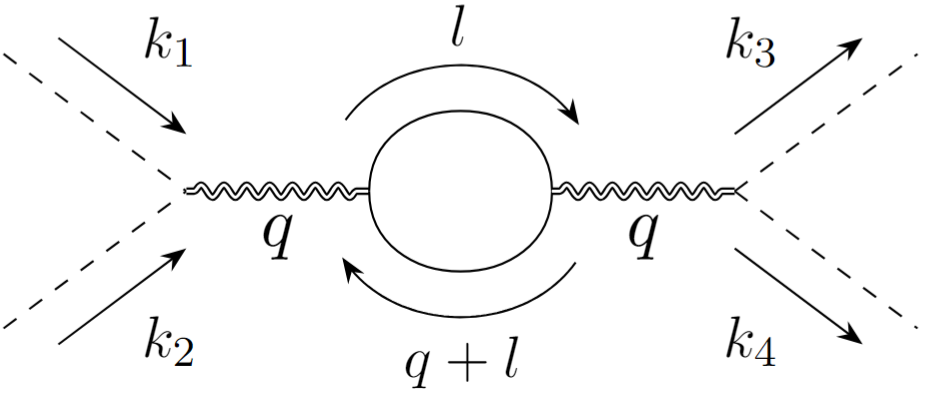}}}
    \label{bubble loop total feynman diagram}
\end{equation}
%\begin{equation}
%    \vcenter{\hbox{$i\mathcal{M}_2$}} \quad=\quad \vcenter{\hbox{\includegraphics[width=0.35\textwidth]{1loopshoestring.png}}}
%    \label{shoestring total feynman diagram}
%\end{equation}
\begin{equation*}
    i\mathcal{M}_2 = \raisebox{-33.5pt}{\includegraphics[width=0.35\textwidth]{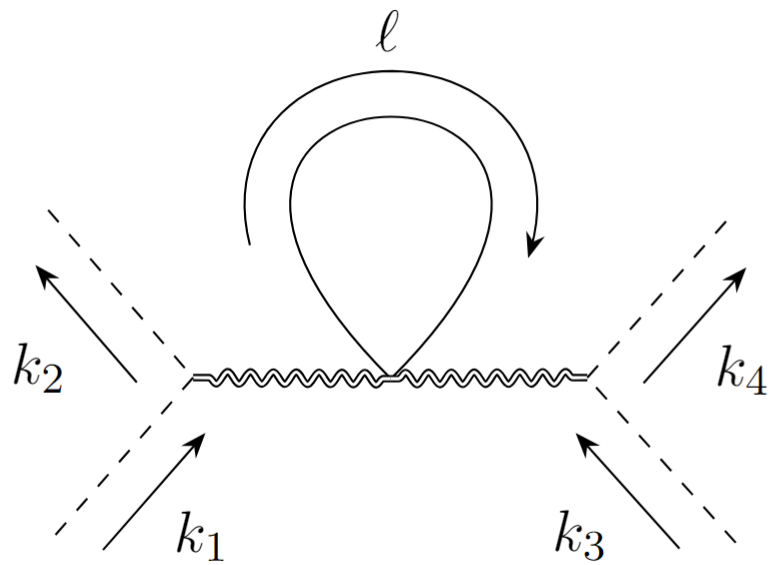}}\label{shoestring total feynman diagram}
\end{equation*}
The general structure of the loop amplitude we are considering is
\begin{equation}
    i\mathcal{M}(q) = \sum_{n=1}^3s_j\,\tau_{\rho\sigma}^{(n)}(k_1,k_2;\m_1)\frac{i\mathcal{P}^{\rho\sigma\lambda\xi}}{q^2 + i\epsilon}\Pi_{\lambda\xi\mu\nu,n}(q)\frac{i\mathcal{P}^{\mu\nu\gamma\delta}}{q^2 + i\epsilon}\tau_{\gamma\delta}^{(n)}(k_3,k_4;\m_2)
    \label{general amplitude structure for diagrams}
\end{equation}
where $s$ denotes the symmetry factor for each diagram. 
The sum is over the 3 diagrams labelled $n=1,2,3,4$, respectively. 

We shall label the mass of the scalar in the loop as $m$, with the understanding that when generalizing to $N$ scalars we simply replace $m\to m_i$ and sum over $i=1,\ldots,N$ at the last step.
Equation \eqref{bubble loop total feynman diagram} is
\begin{align}
    \Pi_{\lambda\xi\mu\nu,1}(q) &= -\frac{1}{2}\mu^{4-d}\int\frac{d^dl}{(2\pi)^d}\frac{\tau_{\lambda\xi}^{(1)}(l+q,l;m)\tau_{\mu\nu}^{(1)}(l,l+q;m)}{(l^2-m^2)((l+q)^2 - m^2)}\nonumber\\
    & = -\frac{1}{2}\mu^{4-d}\int\frac{d^dp}{(2\pi)^d}\frac{\tau_{\lambda\xi}^{(1)}(p+q,p;m)\tau_{\mu\nu}^{(1)}(p,p+q;m)}{(p^2-\Delta)^2}
    \label{matrix element 1pi for bubble diagam}
\end{align}
where $\Delta = m^2-x(1-x)q^2$ and $\mu$ is the scale at which we define the couplings, and we have suppressed the $+i0^+$ in the Feynman propagator for the sake of brevity. The 1PI changes for equation \eqref{shoestring total feynman diagram} as
\begin{equation}
    \Pi_{\lambda\xi\mu\nu,2}(q) = \mu^{4-d}i\int\frac{d^dl}{(2\pi)^d}\frac{\tau^{(2)}_{\lambda\xi\mu\nu}{(l;m)}}{l^2-m^2}%+i\epsilon}
    \label{matrix element 1pi for shoestring}
\end{equation}
where the 4-point vertex between two gravitons and two scalars is given in Appendix \ref{feynman rules appendix section}, equation \eqref{4 point graviton scalar feynman diagram}. We follow the typical procedure for handling these types of diagrams, 
utilizing the $(AB)^{-1}$ Feynman trick, and so on. The numerous amount of terms after contracting each 1PI with the external legs can be sorted into Passarino-Veltman coefficients that can in turn be evaluated from known integrals \cite{Passarino:1978jh}. In the small $\epsilon$-expansion of the integrals with $d=4-2\epsilon$, we keep the divergent $\mathcal{O}(\epsilon^{-1})$ terms and the finite $\mathcal{O}(1)$ terms. 

The expression simplifies in the non-relativistic (NR) limit of the external particles, which we focus on in order to obtain the correction to the Newtonian potential. %Ine momentum of the graviton is $q^\mu = (0,\mathbf{q})$, while 
The external leg momenta becomes $k_{1,2}^\mu = (\m_1,\boldsymbol{0})$ and $k_{3,4}^\mu = (\m_2,\boldsymbol{0})$, and for these loops the exchange graviton's momentum is $q^\mu = (0,\mathbf{q})$. 
This simplifies the expressions since the inner products become
\begin{align}
    k_{1,2}\cdot k_{1,2} = \m_1^2,\:\:\:\:k_{3,4}\cdot k_{3,4}  = \m_2^2,\:\:\:\: q\cdot q = -\mathbf{q}^2,\:\:\:\: k_{1,2}\cdot k_{3,4} = \m_1\m_2.
    \label{R to NR 4 momentum relations}
\end{align}

\subsubsection{One-Loop Diagrams}\label{one loop diagram sections}

Equation \eqref{shoestring total feynman diagram} has a relatively simple tensor structure; we report on that  first. The Feynman diagram of equation \eqref{shoestring total feynman diagram} is given as
\begin{equation}
    i\mathcal{M}_2^{\mu\nu\alpha\beta}(q) = \frac{i\pi^2\kappa^2m^2A_0(m^2)}{2d}\left(\eta^{\alpha\beta}\eta^{\mu\nu}-\eta^{\alpha\nu}\eta^{\beta\mu} - \eta^{\alpha\mu}\eta^{\beta\nu}\right)\label{evaluation of shoestring 1pi matrix element}
\end{equation}
where $A_0(m^2)$ is the tadpole loop integral Passarino-Veltman coefficient \cite{Passarino:1978jh},
\begin{align}
    A_0(m^2) = \frac{\mu^{4-d}}{(2\pi)^d}\int \frac{d^dp}{i\pi}\frac{1}{p^2-m^2}\label{A0 passarino-veltman coefficient}
\end{align}
The matrix element is
\begin{equation}
    i\mathcal{M}^{\mu\nu\alpha\beta}_2 = \frac{i\kappa^2m^4}{128\pi^2}\left(\frac{1}{\epsilon} + \log(\frac{4\pi\mu^2 e^{-\gamma}}{m^2}) + \frac{3}{2}\right)\left(\eta^{\alpha\beta}\eta^{\mu\nu}-\eta^{\alpha\nu}\eta^{\beta\mu} - \eta^{\alpha\mu}\eta^{\beta\nu}\right).\label{evaluated shoestring 1pi matrix element}
\end{equation}

The 1PI for equation \eqref{bubble loop total feynman diagram} however is much more cumbersome, but for clarity amongst the literature, we present the full equation in a form that is readily available for computation. The bubble diagram is comprised of two Passarino-Veltman coefficients, $A_0(m^2)$ and $B_0(q^2,m^2,m^2)$ where
\begin{align}
    B_0(p^2,m^2,m^2) = & \frac{\mu^{4-d}}{(2\pi)^d}\int\frac{d^dq}{i\pi^2}\frac{1}{(q^2-m^2)((q+p)^2-m^2)}\label{B0 passarino veltman coefficient}
\end{align}
The bubble-diagram can then be written as 
\begin{align}
    i\mathcal{M}_1^{\mu\nu\alpha\beta} = -\frac{i\pi^2\kappa^2}{32d(d^2-1)q^4}\left(2A_0(m^2)F^{\mu\nu\alpha\beta} + d\,B_0(q^2,m^2,m^2)G^{\mu\nu\alpha\beta}\right)\label{bubble diagram evaluated matrix element integral}
\end{align}
where the tensors $F^{\mu\nu\alpha\beta}$ and $G^{\mu\nu\alpha\beta}$ are defined in Appendix A. The evaluation of the Passarino-Veltman coefficients leads to the solution provided in \cite{Donoghue:2022chi} but only up to $\mathcal{O}(m^4)$. Here, we confirm this result, and include the $\mathcal{O}(m^2)$ and $\mathcal{O}(m^0)$ terms as well. The result is given in the Appendix in Eq.~(\ref{total evaluation of bubble integrals evaluated}).

We now contract with the external legs as defined in the appendix \eqref{3 point scalar to graviton feynman diagram}. In doing so, we also take the NR limit as described in equation \eqref{R to NR 4 momentum relations}. 
The diagram of \eqref{shoestring total feynman diagram} becomes
\begin{align}
    i\mathcal{M}_2(\mathbf{q}) = -\frac{i\kappa^4 \m_1^2\m_2^2m^4}{128\pi^2\mathbf{q}^4}\left(\frac{1}{\epsilon} + \log(\frac{4\pi\mu^2 e^{-\gamma}}{m^2}) + \frac{5}{2}\right).\label{NR limit of shoestring total diagram}
\end{align}
And the diagram in \eqref{bubble loop total feynman diagram} becomes
\begin{align}
    i\mathcal{M}_1(\mathbf{q}) & =  \frac{i\kappa^4\m_1^2\m_2^2}{7680\pi^2\mathbf{q}^4\epsilon}\left(30m^4 + 10m^2\mathbf{q}^2 + 3\mathbf{q}^4\right)\nonumber\\
    &+\frac{i\kappa^4\m_1^2\m_2^2}{115200\pi^2\mathbf{q}^6}\biggm(
    \!15\left(131+30\log\left(\frac{4\pi\mu^2e^{-\gamma}}{ m^2}\right)\right) m^4 \mathbf{q}^2\nonumber\\
&+10\left(49+15\log\left(\frac{4\pi\mu^2e^{-\gamma}}{ m^2}\right)\right) m^2 \mathbf{q}^4\nonumber
+\left(-12+45\log\left(\frac{4\pi\mu^2e^{-\gamma}}{ m^2}\right)\right)  \mathbf{q}^6    
 \nonumber\\
    &-15 \sqrt{4 m^2 \mathbf{q}^2+\mathbf{q}^4} \left(28 m^4+4 m^2 \mathbf{q}^2+3 \mathbf{q}^4\right) \log \left(\frac{\sqrt{4 m^2 \mathbf{q}^2+\mathbf{q}^4}+2 m^2+\mathbf{q}^2}{2 m^2}\right)\biggm).
   \label{NR limit of bubble diagram}
\end{align}
When summing $\mathcal{M}_1(\mathbf{q})$ and $\mathcal{M}_2(\mathbf{q})$ there are divergences of the form $(\epsilon\,\mathbf{q}^4)^{-1}$ and $(\epsilon\,\mathbf{q}^2)^{-1}$ which are non-local as they grow in the infrared. However, in a consistent renormalization scheme, this pair of divergences can be cancelled by a corresponding pair of counter terms; namely a renormalization of the cosmological constant $\Lambda$ and a renormalization in the gravitational coupling $\kappa$, which we discuss in turn.

\subsubsection{Cosmological Constant Counter Term}\label{cosmological constant counter term section}

Let us first consider the cosmological constant, which appears in the action as $S\supset -\int d^4x\sqrt{-g}\,\Lambda_0$, where we have written the cosmological constant now as $\Lambda\to\Lambda_0$ to indicate that this is the bare value. Then consider the following expansion of the determinant of the metric
\begin{equation}
    \sqrt{-g}\,\Lambda_0 = \Lambda_0\left(1 + \frac{\kappa}{2}h + \frac{\kappa^2}{8}\left(h^2 - 2h_\rho{}^\mu h_\mu{}^\rho\right) + \mathcal{O}(\kappa^3)\right)\label{expansion of determinant of the metric}
\end{equation}
where $h\equiv h^\mu{}_\mu$ is traced over with the Minkowski metric.
The term linear in $h$ is a type of tadpole for the graviton. This makes good sense; a non-zero cosmological constant wants to drive the asymptotic space away from Minkowski to de Sitter or anti-de Sitter. However, what we can do is use the tadpole diagram as a way to find $\delta\Lambda$ where $\Lambda_R = \Lambda_0 - \delta\Lambda$ with $\Lambda_R$ being the renormalized, or measured, value of the cosmological constant that we will eventually take to be zero (we could set it to the observed $\Lambda_R\sim 10^{-123}m_{pl}^4$ but that is negligibly small for this discussion and it would break the asymptotic flat space assumption anyhow).

To proceed, consider the following tadpole expansion up to 1-loop, 
\begin{equation}
\vcenter{\hbox{$i\mathcal{M}^{\mu\nu}_{\text{tadpole}}$}}  = \vcenter{\hbox{\includegraphics[width=0.5\textwidth]{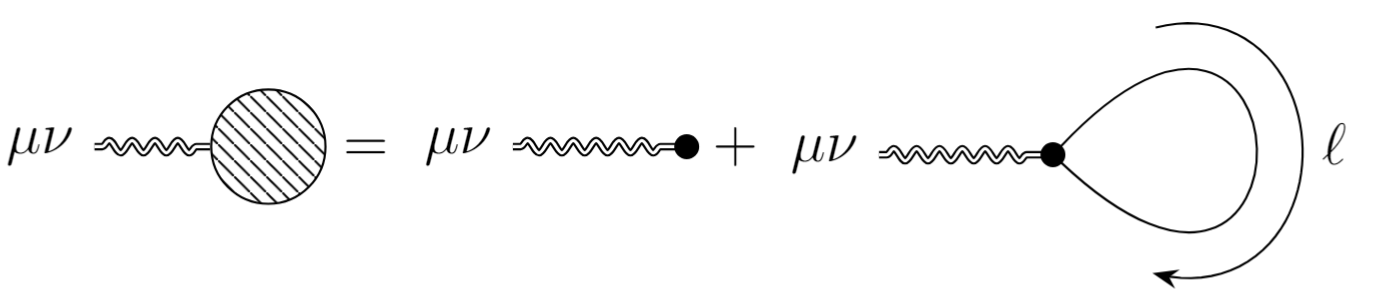}}}
    \label{tadpole diagram total expansion}
\end{equation}

where the first term is provided directly by the above linear term in the action and the second term is a loop generated my a massive scalar. 

For the loop contribution, it is a straightforward calculation from the 3-point vertex of a graviton and two scalars given in equation \eqref{3 point scalar to graviton feynman diagram}  letting all the momentum be the same (while multiplying by the scalar propagator and integrating over all momenta). This calculation yields
\begin{align}
\vcenter{\hbox{$i\mathcal{M}^{\mu\nu}_{\text{tadpole},1}$}} & = \vcenter{\hbox{\includegraphics[width=0.15\textwidth]{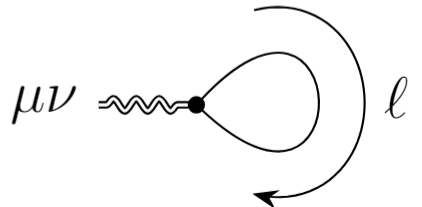}}}\nonumber\\
& = 
    \frac{i}{2}\left(\frac{-i\kappa}{2}\right)\mu^{4-d}\int\frac{d^dl}{(2\pi)^d}\frac{2l^\mu l^\nu - \eta^{\mu\nu}(l^2-m^2)}{l^2-m^2}\nonumber\\
& = \frac{i\kappa m^4 \eta^{\mu\nu}}{128\pi^2}\left(\frac{1}{\epsilon} + \log(\frac{4\pi\mu^2 e^{-\gamma}}{m^2}) + \frac{3}{2}\right).
\label{tadpole matrix elements not traced over}
\end{align}
%from which we can read off $\delta\Lambda$.
Then using the Feynman rule that the linear term in the action gives $-i\kappa\Lambda \eta^{\mu\nu}/2$, equation \eqref{tadpole diagram total expansion} therefore becomes
\begin{equation}
 -   \frac{i\kappa\Lambda_R}{2}\eta^{\mu\nu} = -\frac{i\kappa\Lambda_0}{2}\eta^{\mu\nu} +\frac{i\kappa m^4 \eta^{\mu\nu}}{128\pi^2}\left(\frac{1}{\epsilon} + \log(\frac{4\pi\mu^2 e^{-\gamma}}{m^2}) + \frac{3}{2}\right).\label{algebra and elements of tadpole feynman diagram expansion}
\end{equation}
From this we can read off what the relationship is between the bare and the renormalized cosmological constant at one-loop
\begin{equation}
    \Lambda_0 =\Lambda_R + \frac{m^4}{32\pi^2}\left(\frac{1}{\epsilon} + \log(\frac{4\pi\mu^2 e^{-\gamma}}{m^2}) + \frac{3}{2}\right). \label{renormalized cosmological constant}
\end{equation}

Returning to the perturbed expansion of $\sqrt{-g}$ in equation \eqref{expansion of determinant of the metric}, we can find the Feynman rule for the cosmological constant's contribution to the graviton propagator. After some index manipulation, it is straightforward to show this is

\begin{equation}
    \vcenter{\hbox{\includegraphics[width=0.3\textwidth]{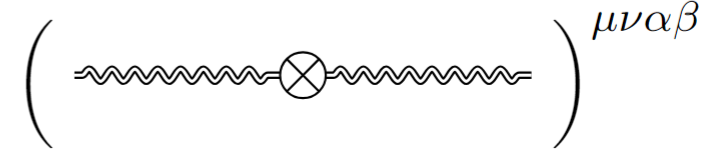}}} \quad =\:\:\vcenter{\hbox{$-\frac{i\Lambda_0\kappa^2}{4}\left(\eta^{\mu\nu}\eta^{\alpha\beta} - \eta^{\alpha\mu}\eta^{\beta\nu} - \eta^{\mu\beta}\eta^{\nu\alpha}\right)$}}
    \label{renormalized feynman diagram graviton propagator}
\end{equation}

This provides a third diagram for the $2\to 2$ scattering of particles, as this generates the following contribution to the amplitude
\begin{equation}
    \vcenter{\hbox{$i\mathcal{M}_\Lambda$}} \quad = \quad \vcenter{\hbox{\includegraphics[width=0.3\textwidth]{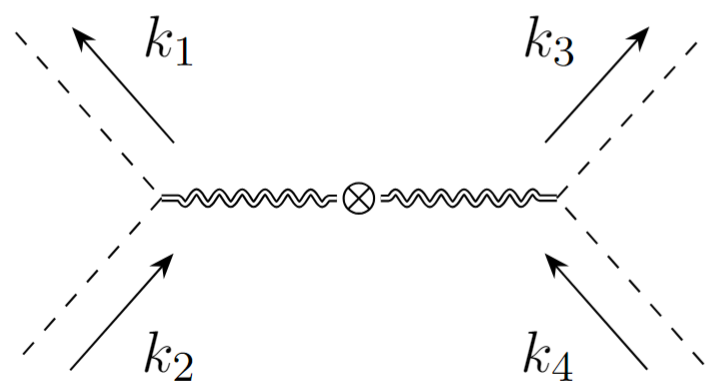}}}
    \label{renorm grav propagator with external legs}
\end{equation}
By inserting the vertices to the external particles and again taking the NR limit, this becomes,
\begin{equation}
    i\mathcal{M}_\Lambda = \frac{i\kappa^4\m_1^2\m_2^2m^4}{256\pi^2\mathbf{q}^4}\left(\frac{1}{\epsilon} + \log(\frac{4\pi\mu^2 e^{-\gamma}}{m^2}) %+ 32\pi^2\Lambda_R  
    +\frac{5}{2}\right),\label{NR limit of cosmological constant total matrix element}
\end{equation}
where we have set the renormalized cosmological constant $\Lambda_R=0$.
The sum of this matrix element $\mathcal{M}_\Lambda$ with the matrix elements $\mathcal{M}_1$ and $\mathcal{M}_2$ cancels the $(\epsilon\mathbf{q}^4)^{-1}$ divergence exactly.

\subsubsection{Newton's Constant Counter Term}\label{Newtons constant counter term section}

Finally, we still have the $(\epsilon\mathbf{q}^2)^{-1}$ divergence, which can be handled by renormalizing Newton's constant $\kappa^2=32\pi G$. In terms of a Feynman diagram, this becomes a counter term to the 3-vertex between scalars and a graviton given as

\begin{equation}
    \vcenter{\hbox{$i\mathcal{M}_{\kappa_0}$}} \quad = \quad \vcenter{\hbox{\includegraphics[width=0.225\textwidth]{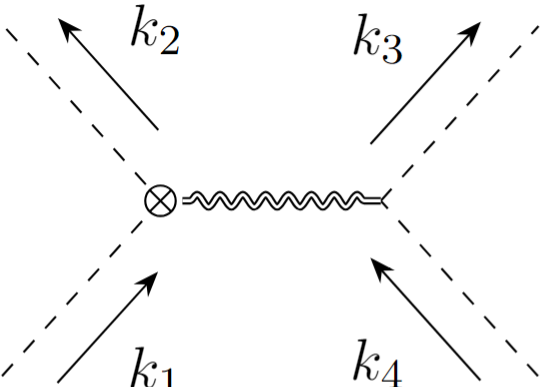}}}
    \label{counter diagram corresponding to renormalized newtons constant}
\end{equation}
This corresponds to writing the coupling constant $\kappa^2\rightarrow \kappa_0^2$ and incorporating a new term in our total 1-loop amplitude (which is just the tree diagram re-labeled)
\begin{equation}
    i\mathcal{M}_\kappa(\mathbf{q}) = \frac{i\kappa_0^2\m_1^2\m_2^2}{2\mathbf{q}^2}. \label{renormalizing AT newtons constant}
\end{equation}
We then require that the sum of this matrix element $\mathcal{M}_\kappa(\mathbf{q})$ and $\mathcal{M}_1(\mathbf{q})$, $\mathcal{M}_2(\mathbf{q})$, $\mathcal{M}_\Lambda(\mathbf{q})$ equates to the observed Newton's constant $\kappa^2=32\pi G$ at the scale of interest. In doing so, we can solve for what $\kappa_0^2$ must be for the total amplitude to be the standard matrix element corresponding to the Newtonian potential
\beq
i\mathcal{M}(\mathbf{q})=\frac{i\kappa^2\m_1^2\m_2^2}{2\mathbf{q}^2}\,\,\,\,\,\mbox{at low momenta $\mathbf{q}\to 0$}
\eeq
(where we mean both the tree level and the counter term.)
We find
\begin{align}
    \kappa_0^2 = \kappa^2 - \frac{m^2\kappa^4}{ 384\pi^2}\left(\frac{1}{\epsilon} +\log(\frac{4\pi\mu^2 e^{-\gamma}}{m^2}) + 2\right). \label{what newton constant must be in NR limit to cancel}
\end{align}
We then insert this back into Eq.~(\ref{renormalizing AT newtons constant}).

\subsubsection{Total Amplitude}\label{total amplitude section}

We can now form the total amplitude by combining the 4 diagrams (including the tree diagram in what we call $\mathcal{M}_\kappa$)
\beq
\mathcal{M}=\mathcal{M}_1+\mathcal{M}_2+\mathcal{M}_\Lambda+\mathcal{M}_\kappa
\eeq
We find that all non-local divergences cancel and we find the following total amplitude
\begin{align}
    i\mathcal{M}(\mathbf{q})= & 
     \frac{i\kappa^2\m_1^2\m_2^2}{2\mathbf{q}^2}
    +\sum_i\biggm[\frac{i\kappa^4\m_1^2\m_2^2}{2560\pi^2\tilde\epsilon} 
    + \frac{i\kappa^4\m_1^2\m_2^2}{115200\pi^2\mathbf{q}^6}\biggm(840 \kappa ^4 m_i^4 \mathbf{q}^2+190 \kappa ^4 m_i^2 \mathbf{q}^4\nonumber\\
    &-15 \sqrt{4 m_i^2 \mathbf{q}^2+\mathbf{q}^4} \left(28 m_i^4+4 m_i^2 \mathbf{q}^2+3 \mathbf{q}^4\right) \log \left(\frac{\sqrt{4 m_i^2 \mathbf{q}^2+\mathbf{q}^4}+2 m_i^2+\mathbf{q}^2}{2 m_i^2}\right)\biggm)\biggm]\label{expanded version of Jq2 for total amplitude}
\end{align}
where any constants were absorbed into a rescaled $\tilde\epsilon$ and we performed a sum over $i=1,\ldots,N$ for each scalar. Notice that in the limit as $\mathbf{q}^2\ll m^2$  Eq.~\eqref{expanded version of Jq2 for total amplitude} becomes
\begin{equation}
    i\mathcal{M}(\mathbf{q}) = \frac{i\kappa^2\m_1^2\m_2^2}{2\mathbf{q}^2} + \sum_i\frac{i\kappa^4\m_1^2\m_2^2}{2560\pi^2\bar{\epsilon}}%\left(\frac{1}{\tilde\epsilon} + \log(\frac{4\pi\mu^2 e^{-\gamma}}{m^2}) -\frac{7}{3}\right) 
    + \mathcal{O}(\mathbf{q}^2)
\end{equation}
(we again  absorbed constants into another rescaled $\epsilon$).
We see that the only non-local piece is the standard tree-level result of Newtonian gravity, and the only divergence is local in that it does not multiply any non-analytic factors of $\mathbf{q}$. This shows that the renormalization procedure not only eliminates any $1/\mathbf{q^2}$ and $1/\mathbf{q}^4$ divergences, but it also renders the low energy physics as standard. The local divergence can be cancelled by the introduction of counter terms $R^2$ and $R_{\mu\nu}R^{\mu\nu}$, but these do not affect long distance physics.

\subsection{Corrections to the Newtonian Potential}\label{corrections to the newtonian potential section}

\begin{figure}[t]
    \centering
    \includegraphics[width = 12cm]{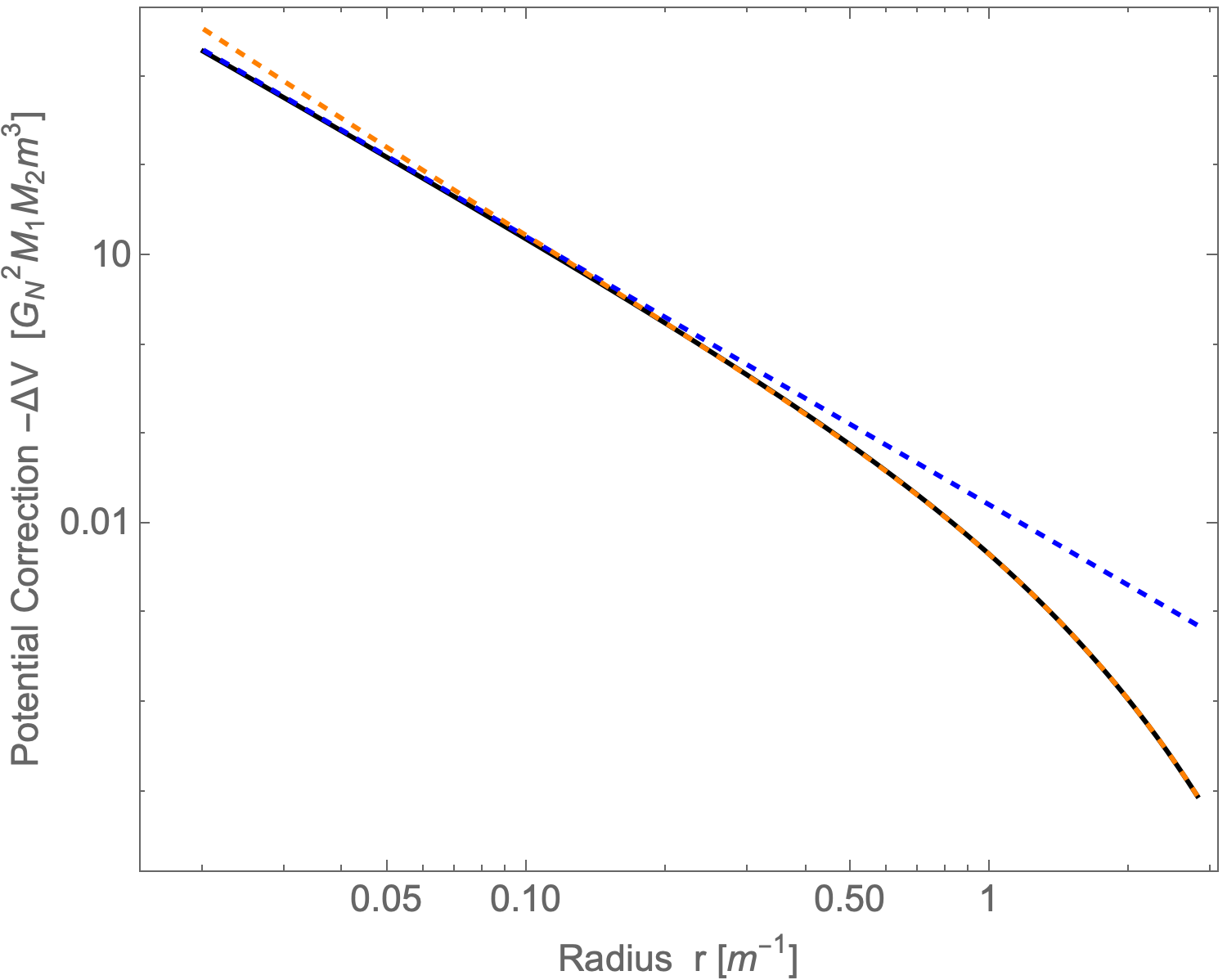}
    \caption{Correction to the potential $\Delta V$ between point masses due to a scalar ($N=1$) running in a loop versus distance $r$. 
    The black curve is the full numerical result from Fourier transforming Eq.~(\ref{expanded version of Jq2 for total amplitude}). The dashed blue curve is the leading term of the small $m\,r$ result given in Eq.~(\ref{massless limit for massive scalar correction newton}). The dashed orange curve is the large $m\,r$ result given in Eq.~(\ref{correct massive limit for massive scalar in a loop correction}).}
    \label{potential}
\end{figure}

We are now in a position to Fourier transform equation \eqref{expanded version of Jq2 for total amplitude} to find what the correction to Newton's potential is when contributions are from a massive scalar in a loop. 
The Fourier transformation from the graviton amplitude to the gravitational potential energy between a pair of point masses is
\beq
    V(\mathbf{r}) = \frac{1}{2\m_1}\frac{1}{2\m_2}\int\frac{d^3q}{(2\pi)^3}\,e^{i\mathbf{q}\cdot\mathbf{r}}\mathcal{M}(\mathbf{q})
    \eeq
Using spherical symmetry, this simplifies to
\beq
V(r)    = \frac{1}{2\m_1}\frac{1}{2\m_2}\frac{1}{ 2\pi^2}\int_0^\infty dq\,q^2\,\frac{\sin(q\,r)}{ q\,r}\mathcal{M}(q).
\eeq
However, an analytic result in terms of elementary functions is not possible due to the presence of logarithms of radicals in Eq.~(\ref{expanded version of Jq2 for total amplitude}). Furthermore, one must be careful with the high $q$ limit as the integral is not absolutely convergent.
The latter can be readily handled by incorporating an appropriate regulator, such as $e^{-\alpha\mathbf{q}}$, for some coefficient $\alpha>0$ that is taken to $0^+$ at the end. We write our answer as
\beq
V(r)=V_N(r)+\Delta V(r)
\eeq
where 
\beq
V_N(r)=-\frac{G\m_1\m_2}{ r}
\eeq
 is the standard Newtonian potential, and $\Delta V$ is the quantum correction. We have carried out the integral numerically, with the result given as the black curve in Figure \ref{potential}.
Apart from performing the integral numerically, we will report on analytical results in two different regimes, namely $mr\ll1$ and $mr\gg 1$.

\subsubsection{Short Distance Regime}\label{short distance regime section}

In the small $mr$ regime, we first expand out equation \eqref{expanded version of Jq2 for total amplitude} for large $q$ and Fourier transform term by term with formulae given in appendix B. We find the following correction to Newton's potential in this regime as
\begin{equation}
    \Delta V(r) = %-\frac{G\m_1\m_2}{r} 
    -N \frac{G^2\m_1\m_2}{20\pi r^3} - \sum_i\frac{G^2m_i^2\m_1\m_2}{6\pi r}\left(\log(m_ir) + \gamma + \frac{1}{3}\right)
    \label{massless limit for massive scalar correction newton}
\end{equation}

If we take the  massless limit we obtain just the single term $\Delta V = -NG^2\m_1\m_2/(20\pi r^3)$; this is given as the dashed blue curve in Figure \ref{potential}.
This confirms Ref.~\cite{Burns:2014bva}, but is in disagreement with Ref.~\cite{Faller:2007sy} by a factor of $1/4$. We note that in Ref.~\cite{Faller:2007sy} there is no symmetry factor of $1/2$ in their bubble diagram, and they did not take into account our tadpole Feynman diagrams and the associated renormalization of the cosmological constant. This may account for the missing factor of $1/4$ from their result. 
For completion, we can include other contributions of massless particles, namely the massless graviton, as well as $N_{1/2}$  massless  spin-$1/2$ fermions, and $N_1$ spin-$1$ bosons, in addition to our $N=N_0$ spinless bosons. 
This gives
\begin{equation}
    \Delta V(r) = -\frac{G^2\m_1\m_2}{\pi\,r^3}\left(\frac{41}{ 10}+\frac{N_0}{ 20}+\frac{2N_{1/2}}{15}+\frac{4N_1}{15}\right)
    %\frac{41\hbar G}{10\pi r^2} + \frac{\hbar G}{45\pi r^2}\left(\frac{9}{4}N_0 + 6N_{1/2} + 12N_1\right)
    \label{all corrections for all particles in the massless limit for newtons potential}
\end{equation}

\subsubsection{Long Distance Regime}\label{long distance regime section}

In the large $mr$ regime, we are pushed towards slightly more involved analyses. We can take anticipate that the structure of the newly Fourier transformed potential should be at asymptotically large $r$ by using the asymptotic series
\begin{equation}
    \Delta V(r) = -\sum_iAG^2\m_1\m_2m_i^3\frac{e^{-a\,m_ir}}{(m_ir)^p}\left(1 + \frac{b_1}{m_ir}+\frac{b_2}{m_i^2r^2}+\ldots\right)
    \label{numerical coefficients for newtons potential}
\end{equation}
where $A,a,p,b_1,b_2,\ldots$ are dimensionless constants to be determined. In order to match onto the above amplitude in (\ref{expanded version of Jq2 for total amplitude}), we use the inverse Fourier theorem (and spherical symmetry) to write
\beq
\Delta\mathcal{M}(q)=2\m_1\,2\m_2\,4\pi\int_0^\infty dr\,r^2\,\frac{\sin(q\,r)}{ q\,r}\,\Delta V(r)
\eeq
(where $\Delta\mathcal{M}$ means the total amplitude minus the tree level result $i\kappa^2\m_1^2\m_2^2/(2q^2)$). We then expand the left hand side in powers of $q$ around $q=0$, and on the right hand side we perform the integral, and then expand the result in powers of $q$. However, we must truncate this at some point. We choose to expand the left hand side out to 4 terms, i.e., $\sim q^2,\,\sim q^4,\,\sim q^6$, and $\sim q^{10}$, and we truncate the expansion on the right hand side at $b_1$. 
We then obtain 4 equations for 4 constants $A,a,p,b_1$. The value of $b_1$ is not expected to be accurate at this level of truncation. However, the value of $A,a,p$ is expected to be somewhat accurate. This truncated procedure yields the values:
$A \approx 0.02798$, $a \approx 1.994$, and $p \approx 2.4499$. One could try working to higher order for more precision, however, then the integrals of the terms $b_2/(m_i^2r^2)$ etc. do not converge at small $r$. So this means this procedure cannot achieve full accuracy. 
A more technical treatment involves the method of steepest descent, which yields the slightly altered values of \cite{Frob:2016xte}:
$A=1/(16\sqrt{\pi})$, $a=2$, and $p=2.5$. This gives the potential to leading order in the large $mr$ regime as
\begin{equation}
    \Delta V(r)=-\sum_i\frac{G^2\m_1\m_2m_i^3}{16\sqrt{\pi}\,(m_ir)^{5/2}}e^{-2m_ir}
    %\left[1 - \frac{13}{16mr} + \mathcal{O}((mr)^{-2})\right]\right).
    \label{correct massive limit for massive scalar in a loop correction}
\end{equation}
This is given as the dashed orange curve in Figure \ref{potential}.
We can compare these results to those of \cite{Burns:2014bva} where we now disagree with their result (equation (V.57)) but do agree with \cite{Frob:2016xte}\footnote{In \cite{Frob:2016xte}, they corrected the large mass expansion found in \cite{Burns:2014bva} which may have been due to the neglect of subleading terms in the expansion of special functions used (see section (4.3)).}. However, numerically we do have agreement with \cite{Frob:2016xte} in the massless and large mass limit.

We can also show direct equivalence between our results and theirs in the following manner.
From the Fourier transformation of the amplitude to Newton's potential, we can re-write it as (subtracting off Newton's potential to focus on the correction)
\begin{equation}
    \Delta V(r) = \frac{1}{16\pi^2M_1M_2}\frac{1}{ir}\int_{0}^\infty dq\: q(e^{iqr} - e^{-iqr})\mathcal{M}(q)e^{-\zeta\abs{q}}\label{eq: potential ready for analytic continuation}
\end{equation}
%\acc{Note to self: change the UV-regulator variable to something OTHER than alpha.}
where we inserted a UV-regulator for completeness $e^{-\zeta\abs{q}}$ to handle the divergences, but this will not be our primary focus nor impede our analysis and will drop it for further analysis. Notice that when we analytically continue the momentum $q\rightarrow ik$, our amplitude $\mathcal{M}(ik)$ will have a branch cut due to the logarithmic term between $-2m<k<2m$. Only considering positive mass terms, we can include these terms by computing the discontinuity across the branch cut namely
\begin{equation}
    \Delta V(r) = -\frac{1}{16\pi^2M_1M_2}\frac{1}{r}\int_{2m}^\infty dk\:ke^{-kr}\:\text{Disc}\:\mathcal{M}(ik)\label{eq: Newtons potential in terms of discontinuity}
\end{equation}
where the discontinuity is defined as $\text{Disc}\:F \equiv F(k+i0) - F(k-i0)$. Computing the discontinuity of $\mathcal{M}(ik)$ only involves the logarithm (which gives a factor of $\text{Disc}\log(\cdots) = 2\pi i$ when the argument is less than $0$, which occurs when $k>2m$). This gives us the following
\begin{equation}
    \text{Disc}\:\mathcal{M}(ik) = \sum_i\frac{\kappa^4M_1M_2}{3840\pi}\frac{\sqrt{k^2-4m_i^2}}{k^5}\left(28m_i^4 - 4m_i^2k^2 + 3k^4\right)\label{eq: disc of newton amplitude}.
\end{equation}
Therefore, \eqref{eq: Newtons potential in terms of discontinuity} takes the following form
\begin{equation}
    \Delta V(r) = -\sum_i\frac{\kappa^4M_1M_2}{61440\pi^3}\frac{1}{r}\int_{2m}^\infty dk\:e^{-kr}\sqrt{k^2-4m_i^2}\left(3 - \frac{4m_i^2}{k^2} + \frac{28m_i^4}{k^4}\right)\label{eq: newtons potential with eval disc}.
\end{equation}
Here \eqref{eq: newtons potential with eval disc} matches that found in \cite{Burns:2014bva}, and that of \cite{Frob:2016xte} (their eq.~(4.18) that was used to find the massless and massive limits). 
%%FIXED\mh{When carrying out this integral, it differs from eq.(\ref{correct massive limit for massive scalar in a loop correction}) by a factor of 1/2; so something is off. It seems we should not correct [27] by a 1/2.}

\subsection{Comparison to Solar System Bounds}\label{comparison to solar system bounds section}

We can relate our 1-loop prediction to previously known experimental bounds by comparing our $r^{-2}$ coefficient to a potential of the form
\begin{equation}
    V(r) = -\frac{G\m_1\m_2}{r}\left(1 + \alpha\, e^{-r/\lambda}\right)
    \label{experimental newtons potential}
\end{equation}
where $\alpha$ is a dimensionless strength parameter and $\lambda$ is a length scale or range. Multiple experimental techniques have been used to place a bound on corrections to Newton's potential. These are summarized in Figure \ref{experimental bounds on r2 newtons corrections}, which is taken from Ref.~\cite{Adelberger:2003zx}.

Let's compare this analysis to the massive correction given in \eqref{correct massive limit for massive scalar in a loop correction} which was found in the limit that $mr\gg 1$. Although this form is not quite the same, since $\alpha$ in the above figure is taken as a constant, we can generalize this. However, to be precise, we need to specify a distribution for the masses, as we now explore.

\subsubsection{All Masses Equal}

\begin{figure}[t]
    \centering    \includegraphics[width=0.5\linewidth,height=0.5\linewidth]{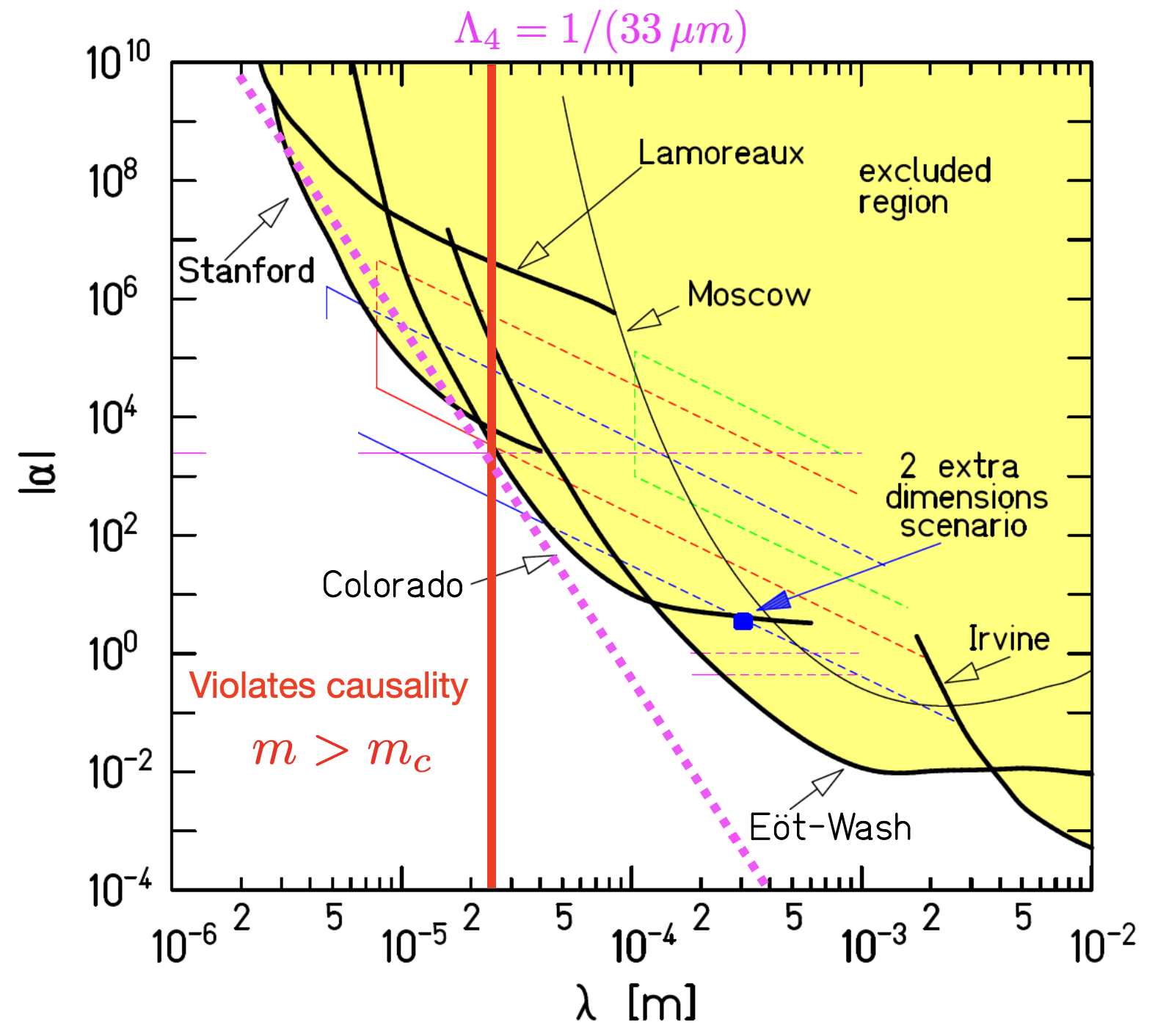}   \includegraphics[width=0.49\linewidth,height=0.49\linewidth]{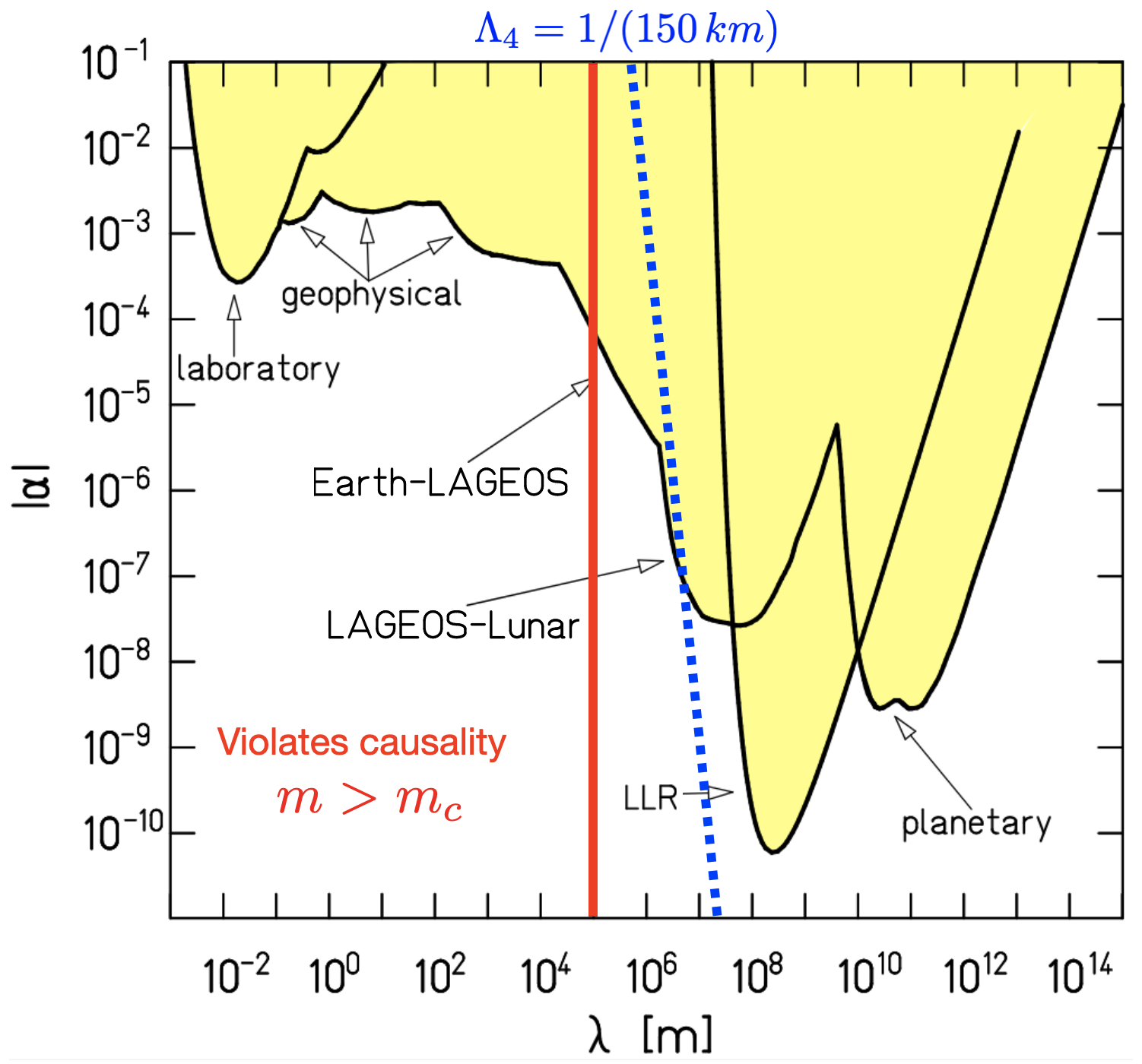}
    \caption{Experimental bounds on the size of corrections to Newton's potential $\alpha$ as a function of scale $\lambda$; taken from Ref.~\cite{Adelberger:2003zx} and \cite{Fischbach:1992fa}. Left side focuses on microscopic scales 
    $10^{-6}m<\lambda<10^{-2}m$,
    while the right side focuses on macroscopic scales 
    $10^{-2}m<\lambda<10^{14}m$.}
    \label{experimental bounds on r2 newtons corrections}
\end{figure}

Let us begin by setting all the masses equal $m=\mn=m_1=\ldots=m_N$. Then we can define an effective value for $\alpha$ from comparing to Eq.~(\ref{correct massive limit for massive scalar in a loop correction}) as
\beq
\alpha\to\alpha(r)=\frac{N\,G\,m^{1/2}}{ 16\sqrt{\pi}\,r^{3/2}}
\eeq
and we identify the scale $\lambda$ as
\beq
\lambda=\frac{1}{2m}.
\eeq
In order to make a comparison we shall evaluate $\alpha$ at $r=\lambda$, which is the regime in which the observations are most sensitive. For example, for the lunar laser ranging (LLR), one can see in Figure \ref{experimental bounds on r2 newtons corrections} that the constraints are most sensitive at around $\lambda=$\,few$\times 10^8$\,m, which is indeed the earth-moon distance. With this choice of $r=\lambda=1/(2m)$, then $\alpha$ is
\begin{equation}
    \alpha_\lambda = \frac{N\,G\,m^2}{\sqrt{32\pi}}.
\end{equation}
%We can scale the correction by a factor of $N$ denoting the number of particles present in the loop. 
We can express this in terms of the previously defined $m_c$ and $N_c$ from Eqs.~(\ref{massbd}) and (\ref{causality bound on N}) as
\begin{equation}
\alpha_\lambda\simeq 2500\left(\frac{m}{m_c}\right)^2\left(\frac{N}{N_c}\right)=2500\left(\frac{m}{m_c}\right)^{6}
%,\:\:\mbox{at current LIGO-Virgo bound $\Lambda_4=1/(150\,\text{km})$}
\label{alpha at current ligo bound}.
\end{equation}
Alternatively, this can be re-written as
\beq
\alpha_\lambda
=2500\left(\frac{m}{ m_0}\right)^6\left(\frac{\Lambda_{4,0}}{ \Lambda_4}\right)^6=2500\left(\frac{\lambda_0}{ \lambda}\right)^6\left(\frac{\Lambda_{4,0}}{ \Lambda_4}\right)^6
\label{alphalamdelta}\eeq
where $m_0\equiv 9.5\times 10^{-13}$\,eV and $\lambda_0\equiv1/(2m_0)=105\:\text{km}$.

Let us insert the current bound from LIGO/Virgo observations, $\Lambda_4=\Lambda_{4,0}=1/(150\,km)$, into this. We obtain the dashed blue curve of Figure \ref{experimental bounds on r2 newtons corrections} right side. The vertical red line is the boundary for causality; to the right of it we have $m<m_c$ and physics can be causal. We see that for the range displayed, this is not a problem. However, we see that we are largely in the yellow region, ruled out by solar system tests. To satisfy the bounds we need
\beq
\lambda \gtrsim 10^4\,\mbox{km}\,\,\,\,\,\,\mbox{for}\,\,\Lambda_4=1/(150\,\mbox{km})
\eeq
This means the cut off on the theory obeys $1/m\gtrsim 2\times 10^{4}\,$km, which is much larger than the size of the black holes that LIGO/Virgo are studying. Hence such an effective theory could not be used in this domain. This would compromise the analysis of Ref.~\cite{Sennett:2019bpc}.

On the other hand, in order for the allowed region to be on the same order as the cutoff, we turn to the left side of Figure \ref{experimental bounds on r2 newtons corrections}. Here we have increased $\Lambda_4$ very significantly to the value $\Lambda_4=1/(33\,\mu m)$. In this case the condition for causality and to also satisfy the bounds are comparable
\beq
\lambda \gtrsim 20\,\mu m\,\,\,\,\,\,\mbox{for}\,\,\Lambda_4=1/(33\,\mu m)
\eeq
In this case, we could use the effective theory all the way to the cut off. But this is not practical, as such extremely small $\Lambda_4$ cannot be realistically obtained with any upcoming measurements, requiring improvement in measurement by about 58 orders of magnitude.

More modestly, if we impose that we are within the solar system bound at the relevant scale for LIGO/Virgo observations, $\lambda\sim 100\,\text{km}$, which from Figure \ref{experimental bounds on r2 newtons corrections} is
\begin{align}
    \alpha_\lambda\simeq 8\times10^{-5},\:\:&\text{at LIGO-Virgo scale}\:\lambda \sim 100\:\text{km}.
\end{align}
Using the above formula for $\alpha_\lambda$ we find that this requires
\beq
%N\lesssim 3\times 10^{-8}\,N_c,\,\,\,\,\,\implies\,\, 
\Lambda_4\gtrsim 1/(8.6\:\text{km})
\eeq
This in turn renders the quartic Riemann curvature corrections to the action to be reduced by over 7 orders of magnitude. This is a difficult challenge to have such high precision, but perhaps an important goal for future interferometers.

%\acc{Here are some comments that could be made to include a general analysis the referee was asking for.}

\subsubsection{Distribution of Masses}

%For a general number of scalars with differing masses, (i.e. $m_1\neq m_2\neq\dots m_i$), we can instead consider a region $(\alpha,\lambda)$ superimposed over the solar system data covered by a generic mass function $\rho(m)$. Namely, consider an effective dimensionless strength parameter 
%\begin{equation}
 %   \alpha_\text{eff}(r) = \int dm\:\rho(m)\alpha_m(r)
%\end{equation}
%with $\lambda(m)  =1/2m$. 

%\acc{But for our analysis, this seems overkill and more of a follow-up paper. This could also just a be a footnote.}

Let us again consider the distributions of masses $\rho(m)$ that we introduced earlier in Eqs.~(\ref{PL}) and (\ref{Exp}). The corrected potential can be changed from the sum in Eq.~(\ref{correct massive limit for massive scalar in a loop correction}) to
\begin{equation}
    \Delta V(r)=-\int dm\,\rho(m)\frac{G^2\m_1\m_2m^3}{16\sqrt{\pi}\,(m r)^{5/2}}e^{-2m r}
    %\left[1 - \frac{13}{16mr} + \mathcal{O}((mr)^{-2})\right]\right).
    \label{intform}
\end{equation}
By carrying out this integral for the power law and exponential distributions, taking the large $r$ limit, and identifying the effective $\alpha$, we obtain
\begin{eqnarray}
&&\alpha\to\alpha_p(r)={N\, G \over 32\sqrt{\pi}\,\mn^{1/2}\,r^{5/2}}
\,\,\,\,\,(\mbox{Power Law})\\
&&\alpha\to\alpha_e(r)=
{N\,G\,\mn^{1/2}\over 32\sqrt{\pi}\,f\,r^{5/2}}
\,\,\,\,\,(\mbox{Exponential})
\end{eqnarray}
where again the scale $\lambda=1/(2\mn)$ in both cases.

As above, we focus on the characteristic  value relvant for observations at $r=\lambda=1/(2\mn)$, where $\alpha$ takes the value
\begin{eqnarray}
    &&\alpha_{\lambda,p}= \frac{N\,G\,\mn^2}{\sqrt{32\pi}}\,\,\,\,\,\mbox{(Power Law)}\\
        &&\alpha_{\lambda,e}= \frac{N\,G\,\mn^3}{\sqrt{32\pi}\,f}\,\,\,\,\,
        \mbox{(Exponential)}
\end{eqnarray}

We now update eq.~(\ref{c3 and c4 coefficient from bern for spin 0}) to a distribution as
\beq
{4\over\Lambda_4^6}= \int dm\,\rho(m)\,\frac{1}{(4\pi)^2}\frac{11}{75600}\frac{1}{m^4}\left(\frac{\kappa}{2}\right)^2
\eeq
(using $\alpha_4=4/\Lambda_4^6$).
We carry out these integrals and solve for $N$ to obtain
\begin{eqnarray}
&&N={96768000\pi^2\mn^4\over 11\kappa^2\Lambda_4^6}
\,\,\,\,\,\mbox{(Power Law)}\\
&&N={58060800\pi^2\mn^3\,f\over 11\kappa^2\Lambda_4^6}
\,\,\,\,\,\mbox{(Exponential)}\label{Nexp}
\end{eqnarray}
where in the exponential case, we have again taken the large $f\gg\mn$ limit, as this is the interesting regime.
Inserting these expressions into $\alpha_\lambda$, and using the above reference values, we obtain an updated version of Eq.~(\ref{alphalamdelta}) as
\begin{eqnarray}
&&\alpha_{\lambda,p}
=12000\left(\frac{m}{ m_0}\right)^6\left(\frac{\Lambda_{4,0}}{ \Lambda_4}\right)^6=12000\left(\frac{\lambda_0}{ \lambda}\right)^6\left(\frac{\Lambda_{4,0}}{ \Lambda_4}\right)^6\,\,\,\,\,\mbox{(Power Law)}\\
&&\alpha_{\lambda,e}
=7300\left(\frac{m}{ m_0}\right)^6\left(\frac{\Lambda_{4,0}}{ \Lambda_4}\right)^6=7300\left(\frac{\lambda_0}{ \lambda}\right)^6\left(\frac{\Lambda_{4,0}}{ \Lambda_4}\right)^6\,\,\,\,\,\mbox{(Exponential)}
\label{alphalamother}
\end{eqnarray}
Note that in the exponential case, the factor of $f$ has dropped out of this result (we have taken the $f\gg\mn$ limit). This may at first sight seem surprising, because in this limit the distribution becomes very broad and so one might expect a smaller value for $\alpha$ as it would seem to be dominated by heavier and heavier particles. However, note that we are fixing $\Lambda_4$ by altering $N$ accordingly. In Eq.~(\ref{Nexp}) $N$ is taken to be proportional to $f$ to ensure there is no change in $\Lambda_4$. This results in the final $\alpha_\lambda$ being independent of $f$ in this limit.

We see that these values for $\alpha_\lambda$ are moderately larger than that for the delta-function distribution in Eq.~(\ref{alphalamdelta}). The pre-factor of 2500 has risen to 12000 and 7300, respectively, in each case. This makes such models all the more ruled out by solar system tests. However, we note that it is only a relatively modest change, since it is only an $\mathcal{O}(1)$ change in pre-factor. So the essential conclusions reached at the end of last subsection carry over.

\subsection{Hawking Radiation}

A concern in these models pertains to the huge number of degrees of freedom $N$ that one would need to integrate out to generate such large corrections. Recall the bound in Eq.~(\ref{causality bound on N}), which indicates we may need to consider $N\sim 10^{82}$ or so (we took $\Lambda_4\sim1/(8.6\,\mbox{km})$ for this estimate). With such a huge number of light particles, one should be concerned if black holes could even exist for a long period of time, rather than rapidly evaporating away via Hawking's process. It is known that a black hole of mass $M$ has a lifetime from the production of $N$ massless (or nearly massless) particles of
\beq
t_{BH}\approx{4\times 10^{67}\over N}\left(M\over M_\odot\right)^3\,\mbox{years}
\label{BHlife}\eeq
where $M_\odot$ is the solar mass.
For moderate $N$ this is extremely long. But for $N\sim 10^{82}$ or so, this becomes a microscopic lifetime. Since we have observed black holes in mergers of masses $M=\mathcal{O}(10)M_\odot$, which must haved lived for a long time, this appears to be ruled out. 

However, the  particles we are studying are not massless. Let us consider the case that they all have the same mass $m$. Then, if we are in a regime in which the black hole temperature is $T_{BH}\lesssim m$, the output radiation is Boltzmann suppressed $\propto e^{-m/T_{BH}}$. The corresponding lifetime is enhanced from Eq.~(\ref{BHlife}) by a factor $\sim e^{m/T_{BH}}$.
Recall that the temperature of a black hole is related to its Schwarzschild radius as
\beq
T_{BH}={1\over 4\pi R_S}={1\over 8\pi G M}.
\eeq
Now the lightest black holes observed (which have the highest temperature) have mass $M\approx 3\,M_\odot$. 
This corresponds to a temperature of 
\beq
T_{BH}\approx 1.7\times 10^{-12}\,\mbox{eV}\left(3\,M_\odot\over M\right).
\eeq
We note that this temperature is comparable to the kinds of masses we have been discussing. However, since it is of the same order, there is not much Boltzmann suppression. 

On the other hand, as we increase precision and increase the value of the masses $m$ that we are sensitive to (which is better for a sensible EFT) then we have $T_{BH}\ll m$ and there is indeed appreciable suppression.
By including the exponential enhancement in Eq.~(\ref{BHlife}), and demanding that the black hole lifetime exceed $5\times 10^9$\,years (as a useful reference), we find the masses would need to be bounded by
\beq
m\gtrsim 1.7\times 10^{-12}\,\mbox{eV}\left(3\,M_\odot\over M\right)\ln\left[10^{-58}N\right]
\eeq
For $N\sim 10^{82}$, this corresponds to $m\gtrsim 9.6\times 10^{-11}$\,eV. Consequently, to be consistent with causality ($\Lambda_4\geq 1.4\,m$) we need $\Lambda_4\geq 1.3 \times 10^{-10}\,\mbox{eV}\approx 1/(1.5\,\mbox{km})$. This is yet another needed improvement in precision.

\section{Conclusions}\label{disccusion and conclusion section}

Motivated by EFT cutoff bounds on higher derivative operators in gravity that was driven by gravitational wave analysis \cite{Sennett:2019bpc}, and causality bounds on the same coefficients \cite{Caron-Huot:2022ugt}, we summarized the bounds on the masses of a range of particles that are integrated out in terms of the couplings in the EFT. To determine the validity of the bound, we proceeded to compute the 1-loop correction to Newton's potential from a massive scalar. This required renormalization of the cosmological constant and Newton's constant which cancelled all non-local divergences. In the non-relativistic limit, this lead to a well defined amplitude whose Fourier transform gives the non-relativistic potential, including corrections to Newton's potential. We carried out this transform numerically for all $r$ and analytically in the small $r$ and large $r$ regimes. This is summarized in Figure \ref{potential}.

We found that if we set the size of the curvature corrections to be controlled by the scale $\Lambda_4 = 1/(150\text{km})$, which is at the limit of current LIGO/Virgo observations, then either: (i) on relevant scales $\lambda\sim 100$\,km, the correction parameter is $\alpha_\lambda\approx2500$ which is ruled out by solar system tests, or (ii) on scales allowed by solar system tests $\lambda\gtrsim 10^4$\,km, the cut off length scale is so large that one cannot use the EFT for merging black holes relevant to LIGO/Virgo. This is summarized in the right hand side of Figure \ref{experimental bounds on r2 newtons corrections}. Finally, we found that if $\Lambda_4\gtrsim1/(8.6\:\text{km})$, then one can test the theory on the relevant LIGO/Virgo scales, but this requires over 7 orders of magnitude improvement in precision. 

We also noted that to avoid over-production of Hawking radiation from black holes, one needs the mass of the scalars to be slightly heavier still. This requires slightly larger values for $\Lambda_4$, and so even further improvement in precision.

A possible extension of our work would be to include other types of massive fields within the loop and determine their contribution to Newton's potential. Importantly, the $\alpha_3$ Wilson coefficient non-zero spin vanishes \cite{Bern:2021ppb}. However, one can then use bounds on the quartic coupling, albeit involving logarithmic corrections.
%Seemingly, having a UV-complete theory with any other massive states that can be integrated out to yield a gravitational EFT will not contribute to cubic Riemann interactions. Then, based on what our causality relation tells us in \ref{mainCbound}, the quartic Wilson coefficient can run off to infinity in strength. But, we would assume that, for example, a 1-loop contribution from a spin-1/2 particle would, after proper renormalization, give a finite effect. This would bring into question the validity of \ref{mainCbound}, and we would recommend using only \eqref{c3 inequality}, \eqref{c4 inequality} which generates logarithmic corrections to the Wilson coefficients. 

\acknowledgments

M. P. H. is supported in part by National Science Foundation grant PHY-2310572.
We thank John Donoghue for helpful discussion.

\appendix
\counterwithin*{equation}{section}
\renewcommand\theequation{\thesection\arabic{equation}}
\section{Feynman Rules and Other Equations}\label{feynman rules appendix section}
\subsection{Feynman Rules}\label{Feynman Rules}
Here we review the ``rules and regulations'' of how to perform and calculate Feynman diagrams for graviton interactions. The majority of this appendix is taken from \cite{Choi:1994ax,Donoghue:1994dn,Donoghue:2017pgk}. As usual, the gravitational Lagrangian is expanded around a flat space background with Minkowski signature $\eta_{\mu\nu} = \text{diag}(+1,-1,-1,-1)$ given as $g_{\mu\nu} = \eta_{\mu\nu} +\kappa\, h_{\mu\nu}$, 
where $\kappa = \sqrt{32\pi G}$ and $h_{\mu\nu}$ is the graviton. From this expansion, along with the canonical scalar field Lagrangian, the Feynman rules can be read off. For a more in-depth procedure on their derivation please consult \cite{Donoghue:1994dn}, otherwise we simply cite the results of the Feynman rules used here. 

The massive scalar propagator is given as

\begin{equation}
    \vcenter{\hbox{\includegraphics[width=0.18\textwidth]{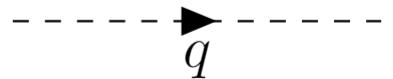}}} \quad = \quad \vcenter{\hbox{$\frac{i}{q^2 - m^2 + i\epsilon}$}}
    \label{scalar propagator}
\end{equation}
where the $i\epsilon$ is the typical Feynman procedure used to ensure analyticty when performing the momentum-loop integrals. The graviton propagator is given as
\begin{equation}
    \vcenter{\hbox{\includegraphics[width=0.18\textwidth]{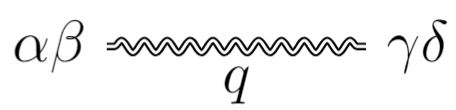}}} \quad = \quad \vcenter{\hbox{$\frac{i\mathcal{P}^{\alpha\beta\gamma\delta}}{q^2 + i\epsilon}$}}
    \label{graviton propagator}
\end{equation}
where 
\begin{equation}
    \mathcal{P}^{\alpha\beta\delta\gamma} = \frac{1}{2}\left(\eta^{\alpha\gamma}\eta^{\beta\gamma} + \eta^{\beta\gamma}\eta^{\alpha\delta}-\eta^{\alpha\beta}\eta^{\gamma\delta}\right).\label{graviton propagator matrix elements}
\end{equation}
The 2-scalar to 1-graviton 3-point vertex is given by
\begin{equation}
    \vcenter{\hbox{\includegraphics[width=0.14\textwidth]{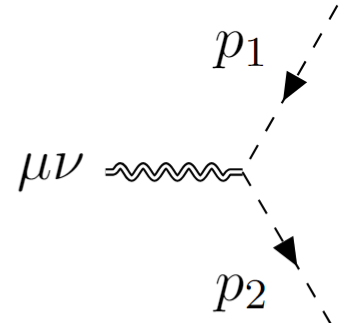}}} \quad = \quad \vcenter{\hbox{$\tau_{(1)}^{\mu\nu}(p_1,p_2;m)$}}
    \label{3 point scalar to graviton feynman diagram}
\end{equation}
where $\tau_1^{\mu\nu}$ is 
\begin{equation}
    \tau_{(1)}^{\mu\nu}(p_1,p_2;m) = -\frac{i\kappa}{2}\left(p_1^\mu p_2^\nu + p_1^\nu p_2^\mu - \eta^{\mu\nu}\left((p_2\cdot p_2)- m^2\right)\right).\label{explicit rule for 3 point feynman diagram}
\end{equation}
There is also the 4-point vertex denoted as $\tau_{(2)}$ given as
\begin{equation}
    \vcenter{\hbox{\includegraphics[width=0.14\textwidth]{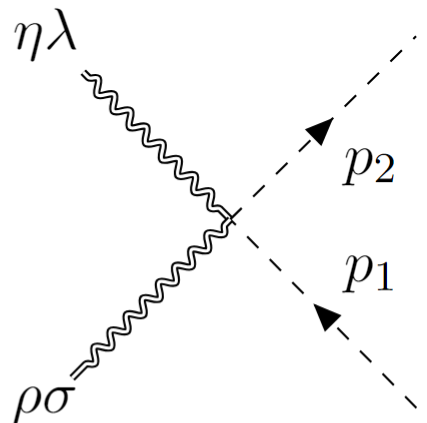}}} \quad = \quad \vcenter{\hbox{$\tau_{(2)}^{\eta\lambda\rho\sigma}(p_1,p_2)$}}
    \label{4 point graviton scalar feynman diagram}
\end{equation}

where $\tau_{(2)}^{\eta\lambda\rho\sigma}$ is
\begin{align}
    \tau_{(2)}^{\eta\lambda\rho\sigma}(p_1,p_2) &  = i\kappa^2\biggm(
    \left(I^{\eta\lambda\alpha\delta}I^{\rho\sigma\beta}{}_{\delta} - \frac{1}{4}\left(\eta^{\eta\lambda}I^{\rho\sigma\alpha\beta} + \eta^{\rho\sigma}I^{\eta\lambda\alpha\beta}\right)\right)(p_\alpha p'_{\beta} + p'_\alpha p_\beta)\nonumber\\
    &-\frac{1}{2}\left(I^{\eta\lambda\rho\sigma} - \frac{1}{2}\eta^{\eta\lambda}\eta^{\rho\sigma}\right)((p'\cdot p)-m^2)\biggm)\label{4 point explicit rule for graviton scalar scattering}
\end{align}
and $I_{\mu\nu\alpha\beta}$ is the ``identity tensor''
\begin{equation}
    I_{\mu\nu\alpha\beta}\equiv\frac{1}{2}\left(\eta_{\mu\alpha}\eta_{\nu\beta} + \eta_{\mu\beta}\eta_{\nu\alpha}\right)\label{identity tensor}.
\end{equation}

\subsection{Other Equations}
\raggedbottom
The definitions of $F^{\mu\nu\alpha\beta}$ and $G^{\mu\nu\alpha\beta}$ that were used to compute the 1PI of the bubble-Feynman diagram are
\begin{align}
    F^{\mu\nu\alpha\beta} = & -2 D \eta^{\beta  \nu } q^{\mu } q^4 q^{\alpha }-2 D \eta^{\beta  \mu } q^{\nu } q^4 q^{\alpha }-2 D^3 q^{\beta } \eta^{\mu  \nu } q^4 q^{\alpha }-{8 D^2 m^2 \eta^{\beta  \nu } q^{\mu } q^2 q^{\alpha }}\nonumber\\
    &-{8 D^2 m^2 \eta^{\beta  \mu } q^{\nu } q^2 q^{\alpha }}-{8 D^2 m^2 q^{\beta } \eta^{\mu  \nu } q^2 q^{\alpha }}-{48 D m^2 q^{\beta } q^{\mu } q^{\nu } q^{\alpha }}-2 D \eta^{\alpha  \nu } q^{\mu } q^4 q^{\beta }\nonumber\\
    &-{8 D^2 m^2 \eta^{\alpha  \nu } q^{\mu } q^2 q^{\beta }}-{8 D^2 m^2 \eta^{\alpha  \mu } q^{\nu } q^2 q^{\beta }}-2 D^3 \eta^{\alpha  \beta } q^{\nu } q^4 q^{\mu }-{8 D^2 m^2 \eta^{\alpha  \beta } q^{\nu } q^2 q^{\mu }}\nonumber\\
    &+4 D \eta^{\mu  \nu } q^{\alpha } q^{\beta } q^4+4 D^2 \eta^{\alpha  \beta } q^{\mu } q^{\nu } q^4+4 D \eta^{\alpha  \beta } q^{\mu } q^{\nu } q^4+16 D^2 \eta^{\alpha  \beta } \eta^{\mu  \nu } m^2 q^4\nonumber\\
    &-8 m^2 \eta^{\alpha  \beta } \eta^{\mu  \nu } q^4+16 D \eta^{\beta  \mu } \eta^{\alpha  \nu } m^2 q^4+8 \eta^{\beta  \mu } \eta^{\alpha  \nu } m^2 q^4+8 \eta^{\alpha  \mu } \eta^{\beta  \nu } m^2 q^4\nonumber\\
    &+{16 D \eta^{\mu  \nu } m^2 q^{\alpha } q^{\beta } q^2}+{16 D \eta^{\beta  \nu } m^2 q^{\alpha } q^{\mu } q^2}+{16 D \eta^{\alpha  \nu } m^2 q^{\beta } q^{\mu } q^2}+{16 D \eta^{\beta  \mu } m^2 q^{\alpha } q^{\nu } q^2}\nonumber\\
    &+{16 D \eta^{\alpha  \mu } m^2 q^{\beta } q^{\nu } q^2}+{2 D^3 q^{\alpha } q^{\beta } q^{\mu } q^{\nu } q^2}+{16 D \eta^{\alpha  \beta } m^2 q^{\mu } q^{\nu } q^2}+{24 D^2 m^2 q^{\alpha } q^{\beta } q^{\mu } q^{\nu }}\nonumber\\
    &+2 D \eta^{\beta  \mu } \eta^{\alpha  \nu } q^4 q^2+2 D \eta^{\alpha  \mu } \eta^{\beta  \nu } q^4 q^2+2 D^3 \eta^{\alpha  \beta } \eta^{\mu  \nu } q^4 q^2-4 D^2 \eta^{\alpha  \beta } \eta^{\mu  \nu } q^4 q^2\nonumber\\
    &-{4 D^2 q^{\beta } q^{\mu } q^{\nu } q^2 q^{\alpha }}-2 D \eta^{\alpha  \mu } q^{\nu } q^4 q^{\beta }+4 D^2 \eta^{\mu  \nu } q^{\alpha } q^{\beta } q^4-16 D m^2 \eta^{\alpha  \beta } \eta^{\mu  \nu } q^4\nonumber\\
    &-4 D \eta^{\alpha  \beta } \eta^{\mu  \nu } q^4 q^2\label{F term for 1pi of bubble diagram}\\
    G^{\mu\nu\alpha\beta} = & -8 D m^2 \eta^{\beta  \nu } q^{\mu } q^4 q^{\alpha }-D^3 q^{\beta } q^{\mu } q^{\nu } q^4 q^{\alpha }-8 D m^2 \eta^{\beta  \mu } q^{\nu } q^4 q^{\alpha }-2 D^2 q^{\beta } \eta^{\mu  \nu } q^2 q^4 q^{\alpha }\nonumber\\
    &-{8 D^2 m^2 q^{\beta } q^{\mu } q^{\nu } q^2 q^{\alpha }}-{48 D m^4 q^{\beta } q^{\mu } q^{\nu } q^{\alpha }}-8 D m^2 \eta^{\alpha  \nu } q^{\mu } q^4 q^{\beta }-8 D m^2 \eta^{\alpha  \mu } q^{\nu } q^4 q^{\beta }\nonumber\\
    &-2 D^2 \eta^{\alpha  \beta } q^{\nu } q^2 q^4 q^{\mu }-2 D \eta^{\alpha  \beta } q^{\nu } q^2 q^4 q^{\mu }+8 D^2 \eta^{\mu  \nu } m^2 q^{\alpha } q^{\beta } q^4-16 D m^4 \eta^{\alpha  \nu } \eta^{\beta  \mu } q^4\nonumber\\
    &+8 D^2 \eta^{\alpha  \beta } m^2 q^{\mu } q^{\nu } q^4-16 D m^4 \eta^{\alpha  \mu } \eta^{\beta  \nu } q^4-16 D m^4 \eta^{\alpha  \beta } \eta^{\mu  \nu } q^4-D \eta^{\beta  \mu } \eta^{\alpha  \nu } q^4 q^4\nonumber\\
    &-D^3 \eta^{\alpha  \beta } \eta^{\mu  \nu } q^4 q^4+2 D^2 \eta^{\alpha  \beta } \eta^{\mu  \nu } q^4 q^4+2 D \eta^{\alpha  \beta } \eta^{\mu  \nu } q^4 q^4+{16 D \eta^{\mu  \nu } m^4 q^{\alpha } q^{\beta } q^2}\nonumber\\
    &+{16 D \eta^{\alpha  \nu } m^4 q^{\beta } q^{\mu } q^2}+{16 D \eta^{\beta  \mu } m^4 q^{\alpha } q^{\nu } q^2}+{16 D \eta^{\alpha  \mu } m^4 q^{\beta } q^{\nu } q^2}+{16 D m^2 q^{\alpha } q^{\beta } q^{\mu } q^{\nu } q^2}\nonumber\\
    &+{16 D \eta^{\alpha  \beta } m^4 q^{\mu } q^{\nu } q^2}+D^3 \eta^{\mu  \nu } q^4 q^{\alpha } q^{\beta } q^2+D \eta^{\beta  \nu } q^4 q^{\alpha } q^{\mu } q^2+D \eta^{\alpha  \nu } q^4 q^{\beta } q^{\mu } q^2\nonumber\\
    &+D \eta^{\alpha  \mu } q^4 q^{\beta } q^{\nu } q^2+D^3 \eta^{\alpha  \beta } q^4 q^{\mu } q^{\nu } q^2-8 D^2 m^2 \eta^{\alpha  \beta } \eta^{\mu  \nu } q^4 q^2+8 D \eta^{\beta  \mu } \eta^{\alpha  \nu } m^2 q^4 q^2\nonumber\\
    &-2 D q^{\beta } \eta^{\mu  \nu } q^2 q^4 q^{\alpha }+2 D^2 q^{\alpha } q^{\beta } q^{\mu } q^{\nu } q^4-D \eta^{\alpha  \mu } \eta^{\beta  \nu } q^4 q^4+{16 D \eta^{\beta  \nu } m^4 q^{\alpha } q^{\mu } q^2}\nonumber\\
    &+8 D \eta^{\alpha  \mu } \eta^{\beta  \nu } m^2 q^4 q^2+D \eta^{\beta  \mu } q^4 q^{\alpha } q^{\nu } q^2
    \label{G term for bubble diagram}
\end{align}
%There is more than likely a much more ``pretty'' form involving Nieuwenhuizen operators.

The uncontracted $\mathcal{M}_1$ is
\begin{align}
    i\mathcal{M}_1^{\mu\nu\alpha\beta}(q) = & \frac{i\kappa^2m^4}{256\pi^2}\left(\frac{1}{\epsilon} + \log(\frac{4\pi\mu^2e^{-\gamma}}{m^2}) + \frac{3}{2} - \frac{8}{5}J(q^2)\right)\left(\eta^{\mu\nu}\eta^{\alpha\beta}+ \eta^{\mu\alpha}\eta^{\nu\beta} + g^{\mu\beta}g^{\nu\alpha}\right)\nonumber\\
    & + \frac{i\kappa^2m^4}{480\pi^2 q^2}J(q^2)\left(q^\alpha q^\beta \eta^{\mu\nu} + \text{all index variations}\right)
     +\frac{i\kappa^2 m^4}{80 \pi^2 q^4}q^\mu q^\nu q^\alpha q^\beta\nonumber\\
    & -\frac{i\kappa^2 m^2}{384\pi ^2}\left(\frac{1}{\epsilon}+\log(\frac{4\pi\mu^2 e^{-\gamma}}{m^2}) + \frac{8}{5}J(q^2) - \frac{109}{15}\right)\biggm(q^\alpha q^\beta \eta^{\mu\nu} + q^\mu q^\nu \eta^{\alpha\beta} - q^2 \eta^{\mu\nu} \eta^{\alpha\beta}\nonumber\\
    &+ \frac{1}{4}\left[q^\alpha q^\mu \eta^{\beta\nu} + q^\beta q^\mu \eta^{\nu\alpha} + q^\alpha q^\nu \eta^{\beta\mu} + q^\beta q^\nu \eta^{\alpha\mu} - q^2 \eta^{\alpha\nu}\eta^{\beta\mu} - q^2 \eta^{\alpha\mu} \eta^{\beta\nu}\right]\biggm)\nonumber\\
    & + \frac{i\kappa^2 m^2}{320\pi^2 q^2}q^\mu q^\nu q^\alpha q^\beta\nonumber\\
    &+\frac{i\kappa^2q^4}{3840\pi^2}\left(\frac{1}{\epsilon}+\log(\frac{4\pi\mu^2 e^{-\gamma}}{m^2}) + J(q^2) + \frac{19663}{3840}\right)\biggm(\eta^{\alpha\nu}\eta^{\beta\mu} + \eta^{\alpha\mu}\eta^{\beta\nu}+\eta^{\alpha\beta} \eta^{\mu\nu}\nonumber\\
    & -\frac{1}{q^2}\left(q^\alpha q^\mu \eta^{\beta\nu} + q^\beta q^\mu \eta^{\alpha\nu} + q^\alpha q^\nu \eta^{\beta\mu} + q^\nu q^\beta \eta^{\alpha\mu}\right)\biggm)\nonumber\\
    &- \frac{i\kappa^2 q^4}{960\pi^2}\left(\frac{1}{\epsilon} + \log(\frac{4\pi\mu^2 e^{-\gamma}}{m^2}) + J(q^2) + \frac{46}{15}\right)\left(\frac{1}{q^2}\left(q^\alpha q^\beta \eta^{\mu\nu} + q^\mu q^\nu \eta^{\alpha\beta}\right)\right)\nonumber\\
    & + \frac{i\kappa^2 q^\alpha q^\beta q^\mu q^\nu}{480\pi^2}\left(\frac{1}{\epsilon} + \log(\frac{4\pi\mu^2 e^{-\gamma}}{m^2}) + J(q^2) + \frac{47}{30}\right).\label{total evaluation of bubble integrals evaluated}
\end{align}
where the $J$ function is defined in the next subsection.

\subsection{Non-local functions \texorpdfstring{$J(q^2)$}{Jq2} and \texorpdfstring{$B(q^2)$}{Bq2}}

following Ref.~\cite{Donoghue:2022chi}, we have defined the function that is denoted as $J(q^2)$ and appears whenn computing the massive-bubble loop in equation \eqref{bubble loop total feynman diagram}
\begin{align}
    J(q^2) & = \int_0^1 dx\: \log(\frac{m^2-x(1-x)q^2}{m^2})\nonumber\\
    & = \frac{1}{q^2}\log(\frac{2m^2 -q^2 + \sqrt{q^2(q^2-4m^2)}}{2m^2})\sqrt{q^2(q^2-4m^2)} -2.
    \label{donoghues non-local function}
\end{align}
However, when building the effective action at quadratic order, as we discuss in a later part of the appendix, we find the following function to be more useful
\begin{align}
    B(q^2) & = \frac{1}{q^2}\log(\frac{2m^2 - q^2 + \sqrt{q^2(q^2-4m^2)}}{2m^2})\sqrt{q^2(q^2-4m^2)} + 2\label{the nonlocal function from 2pt graviton strength EFT action calc}
\end{align}
%where $q^2 = q^\mu q_\mu$. 
The only difference is the sign on the $2$, but this will allow the cancellation of the non-local divergences of the form $(\epsilon q^4)^{-1}$ and $(\epsilon q^2)^{-1}$ to be more manifest.

\subsection{Fourier Transformations}
Here are some Fourier transformations that were useful for finding Newton's potential from the final amplitude \cite{Donoghue:1994dn}. Most are very well known while others require some $\epsilon$-regularization.

In performing the Fourier transforms above, the following integrals are helpful,
\begin{align}
    \int\frac{d^3q}{(2\pi)^3}\frac{e^{i\mathbf{q}\cdot\mathbf{r}}}{\abs{\mathbf{q}^2}} & = \frac{1}{4\pi r}\\
    \int\frac{d^3q}{(2\pi)^3}\frac{e^{i\mathbf{q}\cdot\mathbf{r}}}{\abs{\mathbf{q}}} & = \frac{1}{2\pi^2 r^2}\\
    \int\frac{d^3q}{(2\pi)^3}e^{i\mathbf{q}\cdot\mathbf{r}}\ln(\mathbf{q}^2) & = -\frac{1}{2\pi r^3}.
\end{align}
These are common for massless scalar bubbles inserted into the graviton propagator.

\section{Effective action from the Amplitude}\label{effective action appendix section}

We can put the matrix elements we have calculated into the form of an effective Lagrangian by taking the loop diagrams with the cosmological constant counter-term, and write them as the 1PI of the 2-point graviton propagator. We then trace over with the polarization tensor $\epsilon_{\mu\nu}$. This amounts to summing $\mathcal{M}_\Lambda$, $\mathcal{M}_\kappa$, $\mathcal{M}_1$, and $\mathcal{M}_2$, and contracting with a graviton propagator on each side followed by contracting with two polarization tensors for the remaining gravitons to be placed on-shell. As an example of how this works, we can consider the simplest diagram we have, that of the renormalized cosmological constant of which we found the corresponding feynman diagram counter-term for in equation \eqref{2 pt function of renormalized cosmological constant feynman rule} namely $\mathcal{M}_\Lambda^{\mu\nu\alpha\beta}$. If we contract over with two factors of the graviton propagator, we find the following equation
\begin{equation}
    i\mathcal{M}_\Lambda^{\mu\nu\alpha\beta}(q) = \frac{i\Lambda_0\kappa^2}{4q^4}\left(\eta^{\nu\alpha}\eta^{\mu\beta} + \eta^{\nu\beta}\eta^{\mu\alpha} - \eta^{\mu\nu}\eta^{\alpha\beta}\right).\label{2 pt function of renormalized cosmological constant feynman rule}
\end{equation}
From here, we need to contract with two factors of the polarization tensor which amounts to putting the graviton on-shell. This becomes
\begin{equation}
    i\mathcal{M}_\Lambda(q) = \frac{i\Lambda_0\kappa^2}{4q^4}\left(2\epsilon^{\mu\nu}\epsilon_{\mu\nu}-\epsilon^2\right)\label{putting gravitons on shell for the 2pt function} %this is to get the field strength corresponding to the action
\end{equation}
where we used the fact that the polarization tensor is symmetric. If we then Fourier transform to position space using $h_{\mu\nu} = e^{iq\cdot x}\epsilon_{\mu\nu}$ and $q_\mu\Leftrightarrow i\partial_\mu$ we find
\begin{equation}
    \mathcal{M}_\Lambda=\frac{\Lambda_0\kappa^2}{4\Box^2}\left(2h^{\mu\nu}h_{\mu\nu}-h^2\right).\label{fourier transformed of 2pt matrix elements to real space of cosmological constant}
\end{equation}
This term has a straight forward covariant completion in that if we multiplied the term above by $1 + \frac{\kappa}{2}h+\cdots$ we can see that this will become the following term at the level of the Lagrangian
\begin{equation}
    \mathcal{L}\supset -\sqrt{-g}\Lambda_0\label{action for renormalized cosmo constant from matrix elements}
\end{equation}
where $\Lambda_0$ is given in equation \eqref{renormalized cosmological constant}. The same procedure becomes more complicated when we have to take into account gauge-fixed terms that are found in $\mathcal{M}_1$ and $\mathcal{M}_2$.

To illustrate how this is performed, consider for example the following sum of terms that is simply all the permutations of $\eta^{\mu\nu}q^\alpha q^\beta$ found within the bubble diagram after contracting with two graviton propagators
\begin{align}
    i\mathcal{M}_2(q)^{\mu\nu\alpha\beta}\supset&-\frac{i\kappa^2m^2}{384\pi^2q^4}\frac{1}{\epsilon}\biggm(q^\alpha q^\mu \eta^{\nu\beta} + \eta^{\alpha\nu}q^\beta q^\mu + q^\alpha q^\nu \eta^{\mu\beta} + \eta^{\mu\alpha}q^\beta q^\nu - \eta^{\mu\alpha}\eta^{\nu\beta}q^2\nonumber\\
    &- \eta^{\mu\beta}\eta^{\nu\alpha}q^2 + q^2 \eta^{\mu\nu}\eta^{\alpha\beta}\biggm).\label{example of bubble diagram in 2pt function}
\end{align}
In order to reduce the summed over indices further, we can use the specific gauge we used previously to derive the Feynman rules, namely the harmonic gauge $\partial_\mu h^{\mu\nu} = \frac{1}{2}\partial^\nu h$. It has a Fourier transform of $q_\mu \epsilon^{\mu\nu} = \frac{1}{2}q^\nu \epsilon$. If we substitute this in multiple times into the sum of the terms in \eqref{example of bubble diagram in 2pt function} we find,
\begin{equation}
    i\mathcal{M}_1(q)\supset\frac{i\kappa^2m^2}{368\pi^2 q^4}\frac{1}{\epsilon}\left(h^{\mu\nu}\Box h_{\mu\nu} - 2h\Box h\right).\label{fourier transformed matrix element of bubble diagram to position space}
\end{equation}
In order to covariantly complete of this term in order to find the Lagrangian equivalent does not seem straight forward, but negating factors of $2$ for now, consider the following expansion of $\sqrt{-g}R$ up to $\mathcal{O}(h^2)$ (denoted by $\sim$)
\begin{equation}
    \sqrt{-g}R\sim \frac{\kappa^2}{2}h\left(\partial_\mu\partial_\nu h^{\mu\nu} - \Box h\right).\label{expansion of metric determinant and ricci scalar}
\end{equation}
Recalling that equation \eqref{2 pt function of renormalized cosmological constant feynman rule} was found by tracing over two factors of the graviton propagator (i.e. a factor of $\sim 1/\Box^2$), and that if we take into account the factor of $1/2$ when performing the expansion in equation \eqref{expansion of determinant of the metric}, then the covariant version of equation \eqref{fourier transformed matrix element of bubble diagram to position space} is simply
\begin{equation}
    \mathcal{L}\supset -\frac{\sqrt{-g}}{2(4\pi)^2\epsilon}\frac{m^2}{6}R\label{EFT corresponding to bubble diagram at lagrangian}
\end{equation}
We can do this for the rest of the diagrams, namely, contracting the sum of $\mathcal{M}_\Lambda$, $\mathcal{M}_1$, and $\mathcal{M}_2$ with the renormalized cosmological constant $\Lambda_0$, with two factors of the graviton propagator yields a non-local Lagrangian as well as the already known counter-Lagrangian. First, the \textit{total} counter-Lagrangian that corresponds to renormalized $\kappa$ is \cite{Burns:2014bva, tHooft:1974toh}
\begin{align}
    \mathcal{L}\supset -\frac{\sqrt{-g}}{2(4\pi)^2\overline{\epsilon}}\left(\sum_i\frac{m_i^2}{6}R + \frac{a}{120}R^2 + \frac{b}{60}R_{\mu\nu}R^{\mu\nu}\right)\label{total counter lagrangian}
\end{align}
where $\overline{\epsilon}\equiv 1/\epsilon + \log(4\pi e^{-\gamma})$. For pure gravity, $a=b=1$, but can be corrected when there are $N$ scalars, etc. In deriving the non-local action, this requires more delicacy since there are factors of $1/\Box$ floating around that does not originate from the tracing of the graviton propagators. Using the harmonic gauge that is used in deriving the Feynman rules, $\partial_\mu h^{\mu\nu} = \frac{1}{2}\partial^\nu h$, we can derive the following map between the graviton field and the Ricci tensor
\begin{equation}
    h= -\frac{2R}{\kappa\Box}\:\:\text{and}\:\: h_{\mu\nu}= -\frac{2R_{\mu\nu}}{\kappa\Box}.\label{relation between graviton field and covariant version of ricci and riemann}
\end{equation}
This allows us to find the following non-local action that matches that of what has been found previously in \cite{Donoghue:2022chi,Donoghue:2014yha}, but differs by $3/4$ on the $R^2/\Box^2$ and $R^2\Box$ terms. In fact, any coefficient can be placed on these two terms as long as they have the same value. Having the same value ensures in the limit that $q^2\ll m^2$ the non-local terms cancel. One thing to note before the final answer is that in contracting over with the two graviton propagators, the non-local function $J(q^2)$ becomes $B(q^2)$. The full non-local action is then
\begin{align}
    \mathcal{L} = &\sum_i\biggm(\frac{m^4_i\sqrt{-g}}{40\pi^2}\left(\frac{R_{\mu\nu}}{\Box}B_i(x)\frac{R^{\mu\nu}}{\Box} - \frac{1}{6}\frac{R}{\Box}B_i(x)\frac{R}{\Box}\right) + \frac{m^2_i\sqrt{-g}}{240\pi^2}\left(R_{\mu\nu}\frac{1}{\Box}R^{\mu\nu} - \frac{1}{6}R\frac{1}{\Box}R\right)\nonumber\\
    &+\frac{\sqrt{-g}}{3840\pi^2}\left(RB_i(x)R - 2R_{\mu\nu}B_i(x)R^{\mu\nu}\right)\biggm)\label{total nonlocal action}
\end{align}
where $B_i(x)$ is the Fourier transform of the function $B_i(q^2)$  defined in equation \eqref{the nonlocal function from 2pt graviton strength EFT action calc} for each $m_i$. Now, notice that in the limit $q^2\ll m^2$ (i.e. energy levels above the scalar's mass) then the function $B(x)$ expands as
\begin{align}
    B(x)= -\frac{\Box}{6m^2} + \frac{\Box^2}{60m^4} + \mathcal{O}((\Box/m^2)^3)\label{expansion of B(x) in massive limit}
\end{align}
In this expansion, the non-local components of the Lagrangian $\eqref{total nonlocal action}$ cancel at first order in the expansion of \eqref{expansion of B(x) in massive limit}. This leads to terms on the order of $\Box/m^2$ that can become field re-definitions to the graviton propagator.

\section{Discussion on \texorpdfstring{$R^2$}{R2} and \texorpdfstring{$R_{\mu\nu}R^{\mu\nu}$}{RR}}\label{discussion on R2 appendix section}

We would like to comment on the validity of the total EFT, namely the sum of the action in equations \eqref{the action}, \eqref{total nonlocal action}, and \eqref{action for renormalized cosmo constant from matrix elements}. With the inclusion of the massive scalar in a loop, we were forced to include new terms in the total action that are non-local, but we recover locality in the limit $\Box\ll m^2$ since the $m^2$ and $m^4$ terms cancel at first order. In this limit, the massless non-local contribution becomes local in the sense that the operators now become $\sim\Box R^2/m^2$ which is a massive correction to the graviton's propagator that is suppressed when $m$ is large. In the other limit when $\Box\gg m^2$, the function $B(x)$ behaves as 
\begin{equation}
    B(x) = -\log(\frac{\Box}{m^2}) + 2 + \mathcal{O}(m^2).\label{large momentum limit of B(x)}
\end{equation}
In this limit, we have non-local terms at the level of the Lagrangian which can not be absorbed nor cancelled with any other terms in the non-local Lagrangian. 

In much of the literature when it comes to EFT treatment of gravity, any curvature corrections containing four-derivatives i.e. $R^2$, $R_{\mu\nu}R^{\mu\nu}$, and $(\text{Riem})^2$ are completely neglected. The reasoning for the latter being forgotten is because it can be put into the form of the Gauss-Bonnet term which is a total derivative in the action. The same cannot be said about $R^2$ and $R_{\mu\nu}R^{\mu\nu}$. Instead, as in \cite{Stelle:1977ry}, if we write down an action of the form $S\supset -\int\sqrt{-g}\left(\alpha R_{\mu\nu}R^{\mu\nu} - \beta R^2 - \gamma\kappa^{-2}R\right)$, then the coefficients are $\gamma = 2$ corresponding to Einstein gravity, while $\alpha$ and $\beta$ generate spin-2 and spin-0 massive excitations with masses $m_2 = \sqrt{\gamma/(\alpha\kappa^2)}$ and $m_0 = \sqrt{\gamma/(2(3\beta-\alpha)\kappa^2}$. If we consider the specific case of a point-particle, then the contribution to Newton's potential is given by
\begin{equation}
    V = -\frac{\kappa^2M}{8\pi\gamma r} + \frac{\kappa^2M}{6\pi\gamma}\frac{e^{-m_2r}}{r} - \frac{\kappa^2M}{24\pi\gamma}\frac{e^{-m_0r}}{r}\label{R2 correction to newtons potential from stelle}
\end{equation}
where $M$ is the mass of the point-particle.

However, from viewing $R^2$ corrections as an EFT, it would appear that $R^2$ does not contribute at all as discussed and shown in \cite{AccettulliHuber:2019jqo}. We give a brief derivation of only one of their three different avenues for showing $R^2$ does not contribute to Newton's potential. The argument only requires little group scaling and dimensional analysis.

Consider the 2-to-2 scattering of either scalars or gravitons exchanging a graviton in the background of $R^2$. We would expect the amplitude to be factorizable into the sum of two 3-point amplitudes, namely the 2-to-1 scalar to graviton 3-vertex and the all graviton 3-vertex produced by an insertion of $R^2$. The latter however can not be modified by $R^2$ since all 3-point graviton amplitudes arise from either the expansion of $\sqrt{-g}R$ or $(\text{Riem})^3$. As is known from the spinor helicity formalism, all 3-point vertices are uniquely fixed based on the individual particle's helicity due to little group scaling (for example the general formula can be found in equation (2.99) in \cite{Elvang:2015rqa}\footnote{Or for a quick reference, all 3-point polarization amplitudes can be found using $$\mathcal{A}_3(1^{h_1}2^{h_2}3^{h_3}) = c\langle12\rangle^{h_3-h_1-h_2}\langle 13\rangle^{h_1-h_1-h_3}\langle 23\rangle^{h_1-h_2-h_3}$$}). If we consider the $(1^+2^+3^-)$ amplitude (where the rest of single different helicities can be found in a similar manner) we have two possible helicity structures (one found from the other via conjugation) up to a coupling constant
\begin{equation}
    \mathcal{A}_3(1^+2^+3^-) = c\frac{[12]^6}{[23]^2[13]^2}\:\:\text{or}\:\: \tilde{\mathcal{A}}_3(1^+2^+3^-) = \tilde{c}\frac{\langle 23\rangle^2\langle 13\rangle^2}{\langle 12\rangle ^6}.\label{3 particle amps for only gravitons}
\end{equation}
Since the angle and square brackets each have units of  $m$, then the first 3-point amplitude has units of $m^2$ while the second one has units of $m^{-2}$ being inherently non-local, which we throw away to ensure locality. Thus the first amplitude is the only one we care about, but it can be generated by $R$, and not $R^2$. We can also consider the all-minus (or the all-plus) amplitude $\mathcal{A}(1^\pm 2^\pm 3^\pm)$, or the single graviton scattering with 2 scalars, $\mathcal{A}(1^\pm 2^0 3^0)$. The all-plus amplitudes generates $(\text{Riem})^3$ while the single-plus generates $\phi^2 R$ operators. Both scenarios give angle and square bracket answers with each being local or non-local. 

\begin{comment}
\begin{equation}
    \mathcal{A}(1^-2^-3^-)\sim \langle 12\rangle^2\langle 13\rangle^2\langle 23\rangle^2\:\:\text{or}\:\:\tilde{\mathcal{A}}_3(1^-2^-3^-)\sim \left([12]^2[13]^2[23]^2\right)^{-1}.\label{all minus configurations for 3 particle gravitons}
\end{equation}
Again, the second amplitude is non-local, and thus we will throw it away. The first amplitude however is of the order $m^6$ which can only be generated from a six-derivative term i.e. $(\text{Riem})^3$.
\end{comment}
Notice however that these 3-point amplitudes cannot be produced from a combination of $R\Box R$ such that they can be eliminated from a field re-definition. This takes us back to our non-local Lagrangian in equation \eqref{total nonlocal action}, where if we are working in the limit that $\Box\ll m^2$ where locality is restored, then all of the terms can be eliminated by a re-definition of the graviton field. An immediate push back against this argument however is whether or not one wants to look into non-local physics due to gravitational effects. Then we would suggest working in the limit $\Box\gg m^2$ in which the function $B(x)\sim \log(m/\Box)+2$. The argument from the spinor helicity formalism is then non-applicable as well. 

However, our primary focus is on whether or not the inclusion of $\alpha R^2$ and $\beta R_{\mu\nu}R^{\mu\nu}$ are valid in any EFT of pure gravity. The couplings $\alpha$ and $\beta$ can in fact run \cite{Donoghue:2014yha}, but can still never be found in any scattering experiment. Therefore, based on the arguments above, we would argue that neither $R^2$ nor $R_{\mu\nu}R^{\mu\nu}$ contribute to Newton's potential in any manner when viewing gravitation as an EFT. 

\section{Distribution Functions}\label{AppDistribution}

The function $P_e(y)$ (with $y=\mn/f$) is given by
\beq
P_e(y)={6-6\,y\, \exp(y)\,\Gamma[0,y]\over 2-y+y^2-y^3\, \exp(y)\, \Gamma[0,y]}
\eeq
where $\Gamma$ is the incomplete gamma function $\Gamma[0,y]=\int_y^\infty dt\,e^{-t}/t$.

% Bibliography

%% [A] Recommended: using JHEP.bst file
\bibliographystyle{JHEP}
\bibliography{main}
\end{document}